\documentclass[journal,transmag]{IEEEtran}

\usepackage{amsmath,amssymb,graphicx}
\usepackage{algorithm,algpseudocode,lscape}    
\usepackage{xcolor}
\usepackage{multicol}
\usepackage{mathtools}

\usepackage{pdfpages}

\newcommand{\bx}{\mathbf x}
\newcommand{\bu}{\mathbf u}
\newcommand{\bw}{\mathbf w}
\newcommand{\by}{\mathbf y}
\newcommand{\bv}{\mathbf v}

\newcommand{\tbw}{\tilde{\mathbf w}}

\newtheorem{definition}{\bf Definition}
\newtheorem{remark}{\bf Remark}
\newtheorem{assumption}{\bf Assumption}

\newcommand{\pmat}[1]{\begin{pmatrix} #1 \end{pmatrix}}

\newcommand{\matl}[1]{\begin{matrix*}[l] #1 \end{matrix*}}

\ifCLASSINFOpdf
 
\else
  
\fi

\usepackage[utf8]{inputenc}

\usepackage{tikz}
\usepackage{pgf}
\usepackage{pgfplots}
\pgfplotsset{compat=newest}
\pgfplotsset{plot coordinates/math parser=false}

\pgfplotsset{
    label style={font=\footnotesize},
    xlabel style={yshift=0.0em},
    ylabel style={yshift=-0.0em},
    tick label style={font=\footnotesize },
    label style={font=\footnotesize},
    legend style={font=\footnotesize},
    title style={font=\footnotesize}}
    
    \newlength\fheight
    \newlength\fwidth

\usepackage[numbers]{natbib}
 
\usepackage{caption}
\usepackage{subcaption}
\usepackage{here,float}
\newcommand{\JT}{Joule-Thompson }
\usepackage{psfrag}

\usepackage{transparent}
\ifCLASSINFOpdf
 
\else

\fi

\usepackage[utf8]{inputenc}
\begin{document}

\title{Experimental Investigation of Control Updating Period Monitoring In Industrial PLC-based Fast MPC: \\ Application to The Constrained Control of a Cryogenic Refrigerator}

\author{\IEEEauthorblockN{François Bonne\IEEEauthorrefmark{1},
Mazen Alamir\IEEEauthorrefmark{2} and
Patrick Bonnay\IEEEauthorrefmark{1}}
\IEEEauthorblockA{\IEEEauthorrefmark{1} Univ. Grenoble Alpes, INAC-SBT, F-38000 Grenoble, France CEA, INAC-SBT, F-38000 Grenoble, France}
\IEEEauthorblockA{\IEEEauthorrefmark{2} CNRS, Control Systems Department, University of Grenoble, \\ 11 rue des Math\'{e}matiques, Domaine Universitaire, 38402 Saint-Martin d'Heres, France}
\thanks{This work has been partially supported by the French ANR-Project CryoGreen.}}

% The paper headers
%\markboth{Submitted to IEEE Transactions on Control System Technology}%
%{Shell \MakeLowercase{\textit{et al.}}: Bare Demo of IEEEtran.cls for Journals}
% The only time the second header will appear is for the odd numbered pages
% after the title page when using the twoside option.
% 
% *** Note that you probably will NOT want to include the author's ***
% *** name in the headers of peer review papers.                   ***
% You can use \ifCLASSOPTIONpeerreview for conditional compilation here if
% you desire.

% If you want to put a publisher's ID mark on the page you can do it like
% this:
%\IEEEpubid{0000--0000/00\$00.00~\copyright~2012 IEEE}
% Remember, if you use this you must call \IEEEpubidadjcol in the second
% column for its text to clear the IEEEpubid mark.

% use for special paper notices
%\IEEEspecialpapernotice{(Invited Paper)}

% for Transactions on Magnetics papers, we must declare the abstract and
% index terms PRIOR to the title within the \IEEEtitleabstractindextext
% IEEEtran command as these need to go into the title area created by
% \maketitle.
% As a general rule, do not put math, special symbols or citations
% in the abstract or keywords.
\IEEEtitleabstractindextext{%
\begin{abstract}
In this paper, a complete industrial validation of a recently published scheme for on-line adaptation of the control updating period in Model Predictive Control is proposed. The industrial process that serves in the validation is a cryogenic refrigerator that is used to cool the supra-conductors involved in particle accelerators or experimental nuclear reactors. Two recently predicted features are validated: the first states that it is sometimes better to use less efficient (per iteration) optimizer if the lack of efficiency is over-compensated by an increase in the updating control frequency. The second is that for a given solver, it is worth monitoring the control updating period based on the on-line measured behavior of the cost function. 
\end{abstract}

% Note that keywords are not normally used for peerreview papers.

}

% make the title area
\maketitle

% To allow for easy dual compilation without having to reenter the
% abstract/keywords data, the \IEEEtitleabstractindextext text will
% not be used in maketitle, but will appear (i.e., to be "transported")
% here as \IEEEdisplaynontitleabstractindextext when the compsoc 
% or transmag modes are not selected <OR> if conference mode is selected 
% - because all conference papers position the abstract like regular
% papers do.
\IEEEdisplaynontitleabstractindextext
% \IEEEdisplaynontitleabstractindextext has no effect when using
% compsoc or transmag under a non-conference mode.

% For peer review papers, you can put extra information on the cover
% page as needed:
% \ifCLASSOPTIONpeerreview
% \begin{center} \bfseries EDICS Category: 3-BBND \end{center}
% \fi
%
% For peerreview papers, this IEEEtran command inserts a page break and
% creates the second title. It will be ignored for other modes.
\IEEEpeerreviewmaketitle

\section{Introduction} \label{secintro} 
\noindent Model Predictive Control (MPC) is an attractive control design methodology because it offers a natural way to express optimal objective while handling constraints on both state and control variables \cite{Mayne2000}. MPC design is based on the repetitive on-line solution of finite-horizon open-loop optimal control problems that are parametrized by the state value. Once the optimal sequence of control inputs is obtained, the first control in the sequence is applied over some updating period $\tau_u$ during which, the new problem (based on the next predicted state) is solved and the resulted solution is applied while the prediction horizon is shifted by $\tau_u$ time units and the process is repeated yielding an implicit state feedback. \ \\ \ \\ 
The attractive features of MPC triggered attempts to apply it to increasingly fast systems. For such systems, the need for a high updating rate (small $\tau_u$) may be incompatible with a complete solution of the underlying optimization problem during a single updating period $\tau_u$. This fact fired a rich and still active research area that is shortly referred to by "Fast MPC" (see \cite{Alamir2006,diehl2005real,Zavala2008,Diehl2005IEE} and the reference therein). \ \\ \ \\ Typical issues that are addressed in fast MPC literature concern the derivation of efficient computation of updating steps, reduction of the feedback delay, more or less rigorous computation of the Hessian, etc. Typical proofs of closed-loop stability in that context (see for instance \cite{Diehl2005IEE})  depend on strong assumptions such as the proximity to the optimal solution, the quality of the Hessian matrix estimation, etc. With such assumptions, the corresponding stability proofs take the form of tautological assertions. In other words, when such assumptions are satisfied, the paradigm of fast MPC is less relevant since standard execution of efficient optimizers would anyway give satisfactory results. \ \\ \ \\ 
When the effectively applied control is far from being optimal (which is the case for instance after a sudden change in the set-point value) the hot-start (initialization of the decision variable after horizon shift) it induces for the next horizon does not necessarily decrease the cost function before several iterations. This is because far from the ideal solution, the final stabilizing constraints invoked in the formal proof of \cite{Mayne2000} may be far from being satisfied. On the other hand, if a constant large control updating period is used in order to accommodate for such situations, the overall closed-loop performance would be badly affected.\ \\ \ \\ 
In recent papers \cite{Alamir_pavia2008, Alamir_ECC2013}, investigations have been conducted regarding the impact of the choice of the control updating period $\tau_u$ on the behavior of the cost function. Simple algorithms have been also proposed to monitor on-line the updating period based on the on-line behavior of the cost function to be decreased. More recently \cite{Alamir_ECC2014}, it has been shown that the control updating period choice is intimately linked to the basic iteration being used. The two major facts that come out from these investigations can be summarized as follows: \ \\ \ \\ 
{\bf (Fact 1)} In a constant updating period schemes, it could be interesting to use less efficient (per iteration) algorithms provided that a significantly shorter updating period can be used \cite{Alamir_ECC2014}. This fact enhanced the recent interest \cite{Bemporad2012,Jones2012}  in fast gradient-like algorithms \cite{Nesterov1983} as a simpler approach when compared to second order algorithms. The work in \cite{Alamir_ECC2014} gives a formal explanation for this intuitively accepted fact.  \ \\ \ \\ 
{\bf (Fact 2)} For a given optimization algorithm, the closed-loop performance can be enhanced by an almost computational-free on-line adaptation rule of the control updating period \cite{Alamir_pavia2008, Alamir_ECC2013}. 
\ \\ \ \\ 
Obviously, a combination of the preceding facts holds also, namely, in adaptive frameworks, it can be more efficient to use simpler optimization algorithms provided that the gain induced by a higher updating rate compensates for the lack of efficiency per iteration. \ \\ \ \\
In view of the preceding discussion, the contribution of this paper is twofold: \ \\ \ \\ 
{\bf First contribution.} This paper gives the first industrial validation of the proposed on-line adaptation of the control updating period. The realistic PLC-based implementation framework being used enhances the sensitivity of the closed-loop performance to the adaptation mechanism since it is several orders of magnitude slower than nowadays desk computers. As such, this paper gives a complete and realistic layout to understand the chain of concepts and methods that underline fast MPC paradigm. \ \\ \ \\ 
{\bf Second contribution.}  Although simulation-based assessments have been proposed for Facts 1 and 2 mentioned above, these simulations always used first order gradient-based algorithms. Some promoters of second order algorithms may conjecture that such adaptation would be of no benefit since a second order scheme hardly needs more than a single iteration. This paper invalidates this conjecture by showing that 1) as far as the application at hand is concerned a first-order-like algorithm slightly outperforms a second order algorithm (in the realistic industrial hardware configuration at hand) strengthening Fact 1 in a constant updating period context.  2) the closed-loop performance of this first order algorithm can be improved by on-line adaptation of the control updating period. These two results put together infer that on-line adaptation is worth using even for second order algorithms and that a single iteration is not always sufficient for second order methods in realistic situations.\ \\ \ \\ 
This paper is organized as follows: First, the problem is stated in section \ref{secbackground} by recalling the fast MPC implementation scheme  and the main results of \cite{Alamir_ECC2013,Alamir_ECC2014}. In section \ref{secalgo}, the two algorithms that are used in the validation section are presented which are the {\sc qpOASES} solver \cite{Ferreau2008} and an Ordinary-Differential-Equation (ODE)-based solver that is briefly presented and then applied in the experimental validation.  This second algorithm can be viewed as a first-order algorithm since it is based on the definition of an ODE in which the vector field is linked to the steepest descent direction. In section \ref{secplant}, the process is described, the control problem is stated and the computational PLC used in the implementation of the real-time MPC is presented in order to underline the computational limitation that qualifies the underlying problem as a fast MPC problem.  The main contribution of the paper is given in section \ref{secfastexp}, namely, extensive simulations are first given using the two above cited algorithms and using different constant control updating periods in order to investigate the first fact discussed above. It is in particular shown that for both solvers, the locally (in time) optimal updating period changes dynamically depending on the  context. Moreover, in order to draw conclusions that go beyond the specific case of the PLC at hand, several simulations are conducted using different conjectures regarding the PLC performances. This investigation shows that for the rather performant PLC we actually have today, the first order algorithm gives sightly better results, however, if faster future PLCs were to be used, {\sc qpOASES} would give better results. This is the core message of the paper: the fast MPC paradigm is a matter of combined optimal choices involving the process bandwidth, the optimization algorithm, the available computational device, the control parametrization, etc. Finally, experimental results are shown under adaptive updating period. Section \ref{secconc} concludes the paper and gives hint for further investigation. 
\vskip 0.5cm 
\section{Background} \label{secbackground} 
\subsection{Definitions and Notation}\label{defandnotes}
\noindent Consider a general nonlinear system with state vector $\bx \in \mathbb{R}^{n}$ and a control vector $\bu\in \mathbb{R}^{n_u}$. We consider a basic sampling period $\tau>0$ used to define the piece-wise constant (pwc) control profiles (a sequence of control values in $ \mathbb{R}^{n_u}$ that are maintained constant during $\tau$ time units). As far as the general presentation of concepts is concerned, the general control parametrization is adopted according to which the whole control sequence is defined by a vector of decision variables $p\in \mathbb{R}^{n_p}$ by:
\begin{eqnarray}
\mathcal U_{pwc}(p):= \begin{pmatrix}
u^{(1)}(p)&\dots& u^{(N)}(p)
\end{pmatrix} \in \mathbb U\subset \mathbb{R}^{Nn_u} \label{uparam} 
\end{eqnarray} 
where $u^{(i)}(p)\in \mathbb{R}^{n_u} $ is the control to be applied during the future $i$-th sampling period of length $\tau$ while $\mathbb U\subset \mathbb{R}^{n_u}$ is some admissible set. At this stage, no specific form is required for the system equations describing the dynamic model. The state $\bx_{k+j}$ that is reached - according to the model - after $j$ sampling periods, starting from some initial state $\bx_k$, under the sequence of control inputs $\mathcal U_{pwc}(p)$ and some predicted disturbance $\hat\tbw_k \in \mathbb{R}^{N\cdot n_w}$ is given by:
\begin{eqnarray}
\forall j\in \{1,\dots,N\} \quad \bx_{k+j}=:X(j,\bx_k,p,\hat\tbw_k) \label{predictionmodel} 
\end{eqnarray} 
while the real state that is reached in the presence of true disturbances and/or model mismatched $\tbw_k$ (that takes place over the time interval $[k\tau,(N+j)\tau]$) is denoted by 
\begin{eqnarray}
X^r(j,\bx_k,p,\hat\tbw_k,\tbw_k)
\end{eqnarray} 
In the sequel, explicit mentioning of $\tbw$ is sometimes omitted and the real state evolution is simply denoted by $X^r(j,\bx_k,p,\hat\tbw_k)$.\ \\ \ \\ 
It is assumed that an MPC strategy is defined by the following optimization problem that depends on the current state $\bx$ according to:
\begin{eqnarray}
\mathcal P(\bx)\ :\ \min_{p\in \mathbb{P}} J_0(p,\bx)\quad \mbox{\rm under $g(p,\bx)\le 0$}  \label{cost} 
\end{eqnarray} 
where $\mathbb P\subset \mathbb{R}^{n_p}$ is the admissible parameter set, $J_0$ is the cost function to be minimized while $g(p,\bx)\in \mathbb{R}^{n_c}$ defines the set of inequality constraints. \ \\ \ \\ Recall that in ideal MPC, the solution to (\ref{cost}), say $p^{opt}(\bx)$ is used to define the feedback 
\begin{eqnarray}
K_{MPC}(\bx):=u^{(1)}(p^{opt}(\bx)) \label{defdeidMPC} 
\end{eqnarray} 
Indeed, ideal MPC frameworks assume that the optimal solution $p^{opt}(\bx)$ is instantaneously available. In reality, 
the optimization problem $\mathcal P(\bx)$ is solved using an iterative solver that is denoted by:
\begin{eqnarray}
p^{(q)}=\mathcal S^{(q)}(p^{(0)},\bx) \label{defdemathSq} 
\end{eqnarray} 
where $p^{(0)}$ stands for the initial guess while $p^{(q)}$ is the iterate that is delivered after $q$ successive iterations. In the sequel, the term {\em iteration} refers to the unbreakable set of operations (relative to $\mathcal S$) that is necessary to deliver an update of $p$. In other words, if the time needed to perform a single iteration of $\mathcal S$ on a given platform is denoted by $\tau^\mathcal{S}_1>0$, then no update can be given in less than $\tau_1^\mathcal{S}$ time units. Based on this remark, it seems reasonable to adopt updating instants that are separated by multiples of $\tau_1^{\mathcal S}$, namely:
\begin{eqnarray}
t_{k+1}=t_k+q(t_k)\cdot \tau_1^{\mathcal S}\quad \mbox{\rm with} \quad q(t_k)\in \mathbb N
\end{eqnarray} 
where the $t_k$s are the instants where updated values of $p$ can be delivered for use in the feedback control input. Moreover, we assume for simplicity that the basic sampling period $\tau$ involved in the definition of the control parametrization map $\mathcal U(p)$ is precisely $\tau_1^{\mathcal S}$, namely:
\begin{eqnarray}
\tau=\tau_1^{\mathcal S} \label{tgfr5} 
\end{eqnarray} 
Note that thanks to the flexibility of the parametrization, one can define pwc control profiles in which the control is maintained constant over multiples of $\tau_1^{\mathcal S}$ while meeting (\ref{tgfr5}) so that the latter condition is not really restrictive while it simplifies the description of the implementation framework.\ \\ \ \\ 
Using the notation above, the real-life implementation scheme is defined as follows:
\begin{itemize}
\item[(1)] $i\leftarrow 0$, $t_i\leftarrow 0$, some initial parameter vector $p(t_0)$ is chosen. An initial number of iterations $q(t_0)=q_0\le N$ is adopted.
\item[(2)] The first $q(t_i)$ elements of the control sequence $\mathcal U(p(t_i))$ are applied over the time interval $[t_i,t_{i+1}=t_i+q(t_i)\tau]$.
\item[(3)] Meanwhile, the computation unit performs the following tasks during $[t_i,t_{i+1}]$:
\begin{itemize}
\item[(3.1)] Predict the future state $\hat\bx(t_{i+1})$ using the model and under the above mentioned sequence of controls. The time needed to achieve this very short time ahead prediction is assumed to be negligible for simplicity. 
\item[(3.2)] Perform $q(t_i)$ iterations to get $$p(t_{i+1}):=\mathcal S^{(q(t_i))}(p^+(t_i),\hat \bx(t_{i+1}))$$
where the initial guess $p^+(t_i)$ is either equal to $p(t_i)$ [cold start] or equal to some warm start that is derived from $p(t_i)$ by standard translation technique. \label{step32} 
\end{itemize} 
\item[(4)] At the updating instant $t_{i+1}$ compute the number $q(t_{i+1})$ of iterations to be performed during the next updating period $[t_{i+1},t_{i+2}=t_{i+1}+q(t_{i+1})\tau]$ using Algorithm \ref{algupdateq}  that is recalled in section \ref{recalalgo}. As it has been shown in \cite{Alamir_ECC2013} and recalled hereafter, this updating costs no more than a dozen of elementary operations and can therefore be considered as instantaneous.  
\item[(5)] $i\leftarrow i+1$, Goto step (2).
\end{itemize} 
In the next section, the updating rule for $q(t_{i+1})$ invoked in Step (4) of the implementation scheme is recalled. Note that by adapting $q(t_i)$, the control updating period $\tau_u=q(t_i)\cdot \tau$ is adapted. 
\subsection{Adaptation of the control updating period for a given solver $\mathcal S$} \label{recalalgo} 
\noindent The following definition specifies a class of solvers that is invoked in the sequel and for which the adaptation mechanism recalled in this section can be applied:\\ 
\begin{definition} \label{defmonotonic} 
A solver $\mathcal S$ is said to be monotonic w.r.t the cost function $J: \mathbb{R}^{n_p}\times \mathbb{R}^{n}\rightarrow \mathbb{R}$ if for all $\bx$, the iterations defined by  (\ref{defdemathSq}) satisfies:
\begin{eqnarray}
J(p^{(i)},\bx)\le J(p^{(i-1)},\bx) \label{decrease} 
\end{eqnarray} 
for all $i$. This function is called hereafter the augmented cost function. $\hfill \diamondsuit$\\
\end{definition}
Note that $J$ generally differs from $J_0$ involved in (\ref{cost}) because of the presence of constraints. A typical example of such $J$ is given by the norm of the nonlinear function that gathers the Karush-Kuhn-Tucker (KKT) necessary conditions of optimality and when the solver uses a descent approach such as projected gradient or a specific implementation of Sequential Quadratic Programming (SQP) approach with trust region mechanism. Interior point-based algorithm can also enter in this category under certain circumstances in which the map $J$ would be given by the penalized version of $J_0$ involving barrier functions. \\ 
\begin{remark}\label{remalwaysdecrease} 
The conditions of Definition \ref{defmonotonic}  can be relaxed in the following sense: if a solver $\mathcal S$ satisfies the following condition:
\begin{eqnarray}
J(p^{(i+\ell-1)},\bx)\le J(p^{(i-1)},\bx) \label{decrease2} 
\end{eqnarray}   
for some map $J$, then the solver $\mathcal S^{'}$ that is derived from $\mathcal S$ by:
\begin{eqnarray}
\mathcal S^{'}(p,\bx):=S^{(\ell)}(p,\bx)
\end{eqnarray} 
is monotonic in the sense of Definition \ref{defmonotonic} at the price of having a {\em single} iteration that takes $\ell$-times longer than $\mathcal S$, namely $\tau_1^{\mathcal S^{'}}=\ell\cdot \tau_1^{\mathcal S}$. $\hfill \diamondsuit$
\end{remark}
\ \\
The following assumption is needed for the updating algorithm that can be stated as follows:\\ 
\begin{assumption} \label{boundedbelow} 
The solver $\mathcal S$ is monotonic and the corresponding map $J$ [see Definition \ref{defmonotonic}] is bounded below by a strictly positive real $\underline J$, namely:
\begin{eqnarray}
\forall (p,\bx)\quad J(p,\bx)\ge \underline J>0 \label{underlineJ} 
\end{eqnarray} 
\end{assumption}
Note that the last condition (\ref{underlineJ}) can always be satisfied by adding an appropriate positive constant to the original cost. \ \\ \ \\ 
In order to recall the updating algorithm proposed in \cite{Alamir_ECC2013}, the following notations are needed:\ \\ \ \\ 
$J^+_k:=J\bigl(p^+(t_k),\hat\bx(t_{k+1})\bigr)$ \vskip 0.2cm
\begin{minipage}{0.02\textwidth}
\ 
\end{minipage}
\begin{minipage}{0.45\textwidth} 
the cost function value for the initial hot start $p^+(t_k)$ (before any iteration is performed) and based on the predicted state at the future updating instant $t_{k+1}=t_k+q(t_k)\cdot \tau$.
\end{minipage}
\ \\ \ \\ 
$\hat J_{k+1}:=J\bigl(p(t_{k+1}),\hat\bx(t_{k+1})\bigr)$ \vskip 0.2cm
\begin{minipage}{0.02\textwidth}
\ 
\end{minipage}
\begin{minipage}{0.45\textwidth} 
the cost function value for the delivered value $p(t_{k+1})$ (after $q(t_k)$ iterations) and based on the predicted state at the future updating instant $t_{k+1}=t_k+q(t_k)\cdot \tau$.
\end{minipage}
\ \\ \ \\ 
$J_{k+1}:=J\bigl(p(t_{k+1}),\bx(t_{k+1})\bigr)$ \vskip 0.2cm
\begin{minipage}{0.02\textwidth}
\ 
\end{minipage}
\begin{minipage}{0.45\textwidth} 
the effectively obtained cost function value for the delivered value $p(t_{k+1})$  and for the true state $\bx(t_{k+1})$ that is reached at instant $t_{k+1}=t_k+q(t_k)\cdot \tau$.
\end{minipage}
\ \\ \ \\ 
Based on these definitions, it comes out that the decrease of the augmented cost function can be studied by analyzing the behavior of the ratio $J_{k+1}/J_k$ which can be decomposed according to:
\begin{eqnarray}
\dfrac{J_{k+1}}{J_k}=E_k^r(q(t_k))\times D_k^r(q(t_k))
\end{eqnarray} 
where 
\begin{eqnarray}
E_k^r(q(t_k)):=\dfrac{\hat J_{k+1}}{J^+_k}\quad;\quad D_k^r(q(t_k)):=\dfrac{J_{k+1}}{\hat J_{k+1}}\times \dfrac{J_k^+}{J_k}
\end{eqnarray} 
A deep analysis of the above terms shows that $E_k^r(q)$ is linked to the current efficiency of the solver since it represents the ratio between the value of the augmented cost for the same predicted value $\hat \bx(t_{k+1})$ of the state before and after $q(t_k)$ iterations are  performed. The first ratio $J_{k+1}/\hat J_{k+1}$ in $D_k^r$ is $1$ if the model is perfect since it represents the ratio between two values of the augmented function for the same value $p(t_{k+1})$ of the parameter but for two different values $\hat x(t_{k+1})$ and $x(t_{k+1})$. Finally, the ratio $J^+_k/J_k$ is linked to the quality of the hot start since it represents the predicted ratio between two values of the augmented function before and just after the horizon shift. \ \\ \ \\ 
The algorithm proposed in \cite{Alamir_ECC2013} recalled hereafter updates the number of iterations $q(t_{k+1})$ to be performed during the next updating period so that the contraction ratio:
\begin{eqnarray}
K_{k+1}^r(q(t_{k+1})):=E_{k+1}^r(q(t_{k+1}))\times D_{k+1}^r(q(t_{k+1}))
\end{eqnarray} 
is lower than $1$ and if this is achievable, the updating rule tries to minimize the corresponding expected response time $t_r$ of the dynamics which is linked to the ratio $q/\log(K_{k+1}^r(q))$. \ \\ \ \\ 
This leads to the following algorithm \cite{Alamir_ECC2013}:
\begin{algorithm}[H]  
\caption{Updating rule for $q(t_{k+1}^u)$}
\label{algupdateq}     
\begin{algorithmic}[1] 
\State{{\bf Parameters}  $q_{max}\le N$, $\delta\in \{1,\dots,q_{max}\}$}
\State{{\bf Input data} (available after Step (3.2) page \pageref{step32})}
\State{$\quad$ $q=q(t_k)$, $p^{(0)}=p^+(t_k)$, $p^{(i)}=\mathcal S^{(i)}(p^{(0)},\hat \bx(t_{k+1}))$}
\State{$\quad$ $J_k$, $J^+_k$, $\hat J_{k+1}$, $J_{k+1}$}
\State{Compute the following quantities:}
\begin{eqnarray*}
E^r &\leftarrow& \hat J_{k+1}/J^+_k \\
D^r &\leftarrow& (J_{k+1}J^+_k)/(\hat J_{k+1}J_k)\\
K^r &\leftarrow& E^r\times D^r\\
  \dfrac{\Delta D^r}{\Delta q} &\leftarrow& \dfrac{1}{q}\bigl[D^r-1\bigr]\\
 \dfrac{\Delta E^r}{\Delta q} &\leftarrow& \dfrac{J(p^{(q)},\hat\bx(t_{k+1}))-J(p^{(q-1)},\hat \bx(t_{k+1})}{J(p^{(0)},\hat\bx(t_{k+1}))}\\
 \dfrac{\Delta K^r}{\Delta q}&\leftarrow& E^r\cdot \bigl[\dfrac{\Delta D^r}{\Delta q}\bigr]+D^r\cdot \bigl[\dfrac{\Delta E^r}{\Delta q}\bigr]\\
\dfrac{\Delta t_r}{\Delta q}&\leftarrow& \dfrac{-\log(K^r)+\dfrac{q}{K^r}\times \dfrac{\Delta K^r}{\Delta q}}{\bigl[\log(K^r)\bigr]^2}
\end{eqnarray*}
\State{{\bf If} $K^r\ge 1$ {\bf then} $\Gamma\leftarrow \dfrac{\Delta K^r}{\Delta q}$ {\bf else} $\Gamma\leftarrow \dfrac{\Delta t_r}{\Delta q}$}\vskip 0.1cm 
\State{{\bf Output} $q(t_{k+1})\leftarrow \max\bigl\{2,\min\{q_{max},q-\delta\cdot sign(\Gamma)\}\bigr\}$}
\end{algorithmic}
\end{algorithm}
\noindent Roughly speaking, this algorithm implements a step of size $\delta$ in the descent direction defined by the sign of the approximated gradient $\Gamma$. The step is projected into the admissible domain of $q\in \{2,\dots,q_{max}\}$. More details regarding this algorithm are available in \cite{Alamir_ECC2013}. \ \\ \ \\ 
Section \ref{secfastexp} shows the efficiency of the proposed algorithm when applied to a given solver for the PLC-based implementation of MPC to the cryogenic refrigerator. Before this, the next section gives a simple argumentation that underlines a fundamental trade-off between the efficiency (per iteration) of a solver and the basic corresponding unbreakable computation time $\tau_1^{\mathcal S}$. This is done in adaptation-free context in order to decouple the analysis. 
\subsection{Fundamental trade-off in the choice of solvers} \label{fundtradeoff} 
\begin{figure}[H]
\includegraphics[width=0.48\textwidth]{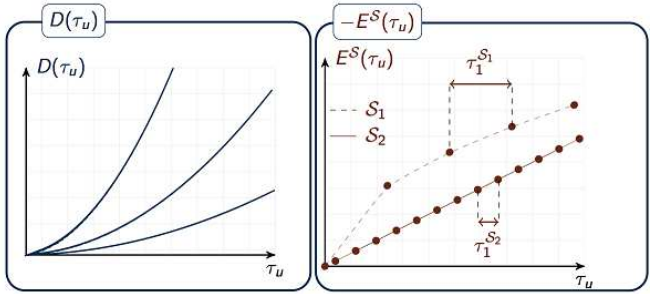}
\caption{Possible allures of the $D_k(\tau_u)$ and $E_k^\mathcal{S}(\tau_u)$ in realistic fast NMPC implementations. The right figure shows the efficiency maps  for two different solvers corresponding to two different computation times per iteration $\tau_1^{\mathcal S_1}$ and  $\tau_1^{\mathcal S_2}$.}\label{compareSf} 
\end{figure}

\noindent Let us consider a solver $\mathcal S$ and the corresponding time $\tau_1^\mathcal{S}$ that is needed to perform the unbreakable amount of computations involved in a single iteration. Given a control updating period $\tau_u$, the number of iterations that can be performed is equal to $q=\lfloor\tau_u/\tau_1^\mathcal{S}\rfloor$ and the corresponding variation of the augmented cost function would be given by:
\begin{eqnarray}
J_{k+1}-J_k:=\underbrace{J_{k+1}-J^+_k}_{-E^\mathcal{S}_k(\tau_u)}+\underbrace{J^+_k-J_k}_{D_k(\tau_u)} \label{DeltaJ} 
\end{eqnarray} 
where here again, $E_k^{\mathcal S}(\tau_u)$ and $D_k(\tau_u)$ are linked to the current efficiency of the solver ($E_k^\mathcal{S}$) and the combined effect of model mismatch and the horizon shift effect on the cost function respectively. Both terms depend obviously on $\tau_u$. Indeed $E_k^\mathcal{S}(\tau_u)$ depends on $\tau_u$ through the number of iterations while $D_k(\tau_u)$ depends on $\tau_u$ since when $\tau_u=0$ then $D_k$ vanishes (no prediction error and no possible bad hot start). Note that $E_k^\mathcal{S}$ and $D_k$ are absolute (non relative) versions of the relative maps $E_k^r$ and $D_k^r$ invoked in section \ref{recalalgo} to introduce Algorithm \ref{algupdateq}. Note also that unlike the efficiency indicator $E_k^\mathcal{S}(\tau_u)$ which heavily depends on the solver, the $D_k(\tau_u)$ term is solver-independent. \ \\ \ \\ 
Figure \ref{compareSf} shows typical allures of these terms for two different solvers $\mathcal S_1$ (most efficient) and $\mathcal S_2$ (less efficient). It can be seen that the iterations of $\mathcal S_1$ are more efficient at the price of longer computation time $\tau_1^{\mathcal S_1}>\tau_1^{\mathcal S_2}$. The dots on the right hand plot recall that the updating can be delivered only at quantized updating instants. 
\noindent Now based on (\ref{DeltaJ}), the decrease of $J_k$ is conditioned by the inequality:
\begin{eqnarray}
E_k^\mathcal{S}(\tau_u)>D_k(\tau_u) \label{ineqtradeoff} 
\end{eqnarray} 
which expresses the need to have the $E_k^\mathcal{S}$ curves above the $D_k$ curve for the adopted value of the updating period.
\begin{figure}[H]
\begin{minipage}{0.49\textwidth}
\framebox{\includegraphics[width=0.45\textwidth]{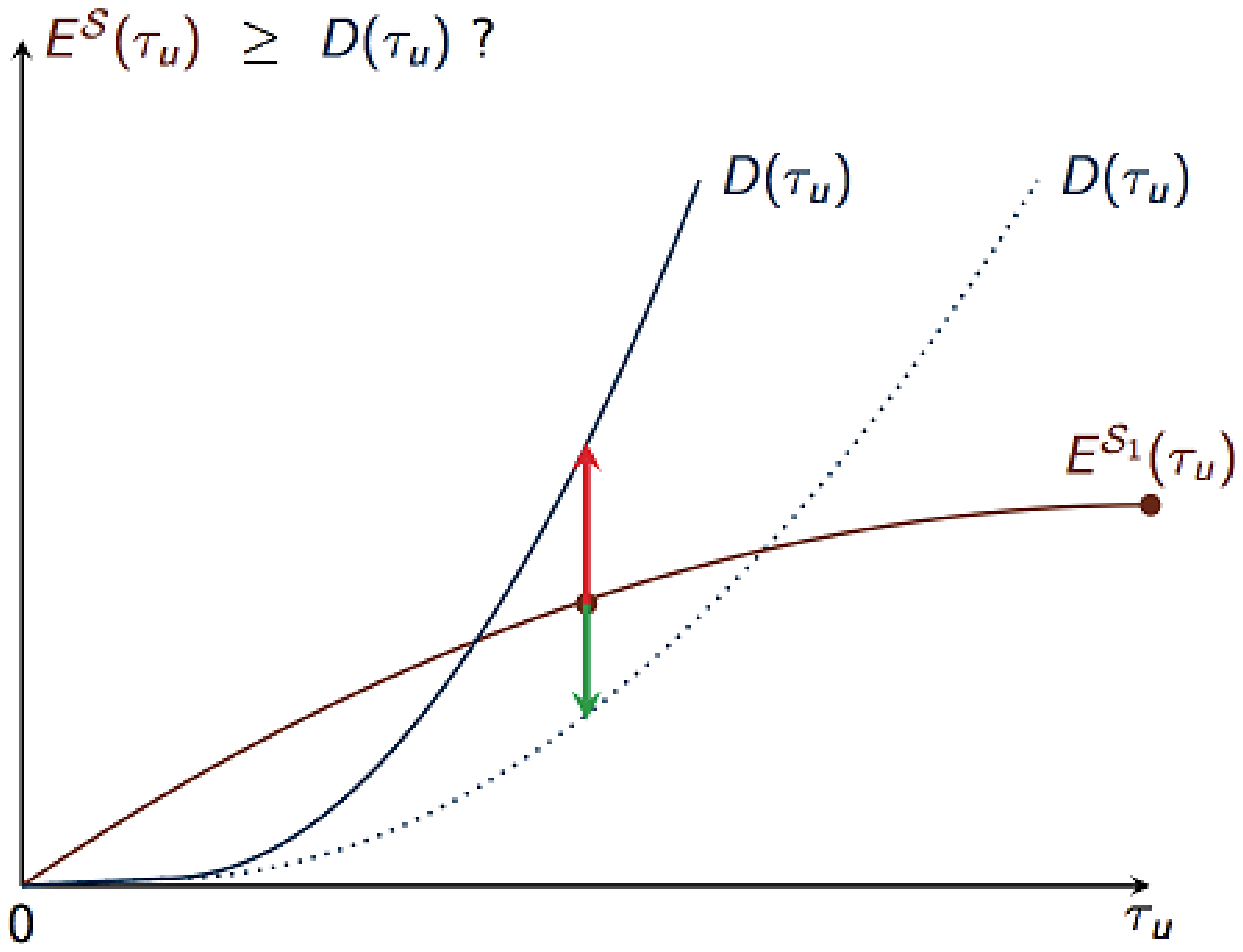}}
\framebox{\includegraphics[width=0.452\textwidth]{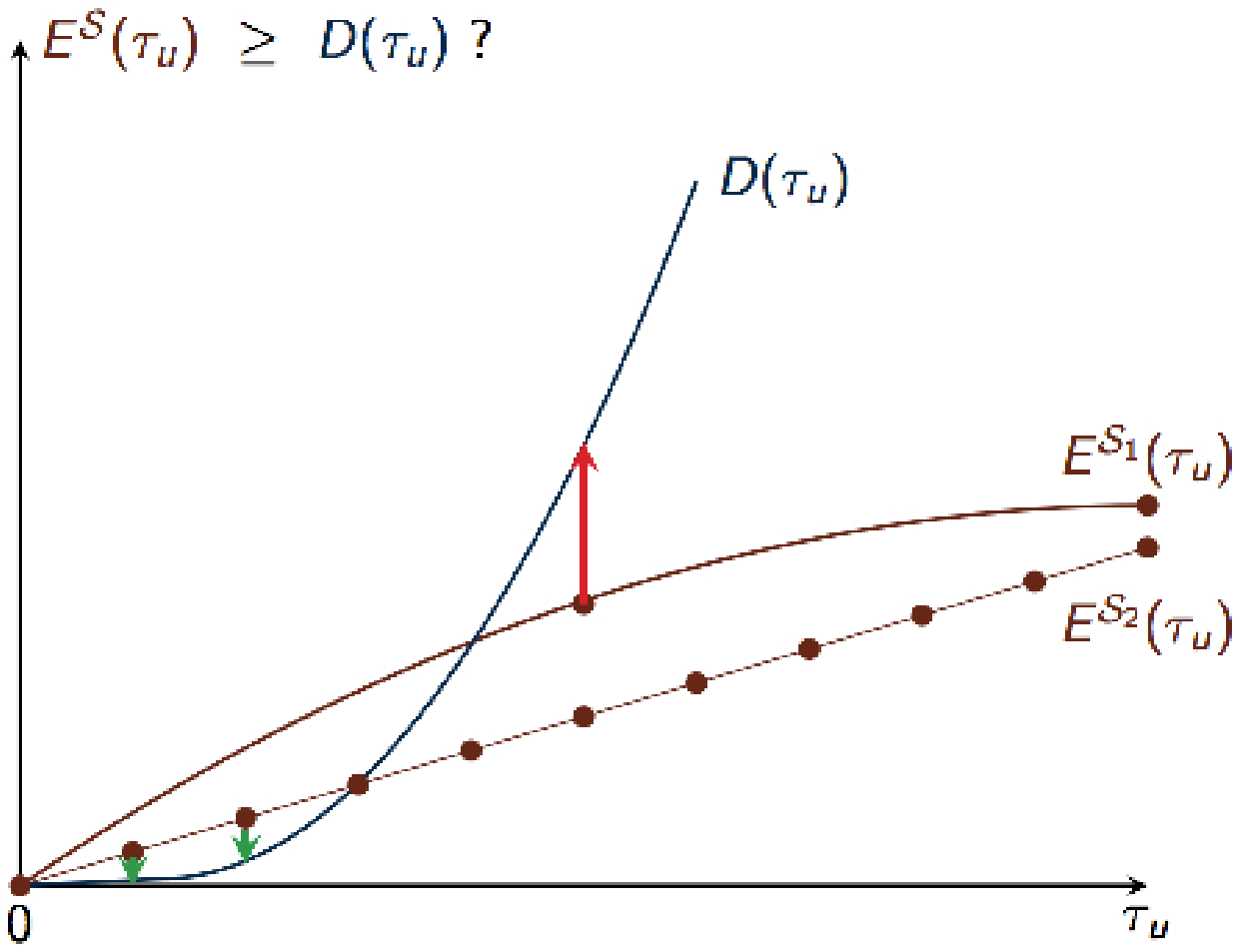}}
\end{minipage}
\caption{(Left) Use of the most efficient solver $\mathcal S_1$: depending on the context, there are possible configurations of $D$ that make the decrease of the augmented cost impossible. (Right) In such cases, the use of the less efficient solver $\mathcal S_2$ enables to decrease the augmented cost thanks to shorter updating periods.} \label{twocases} 
\end{figure}
\noindent Figure \ref{twocases} gives a qualitative illustration of the resulting fundamental trade-off: The left plots shows situations where the use of the more efficient solver $\mathcal S_1$ makes (\ref{ineqtradeoff}) impossible to satisfy whatever is the updating period being used. In such cases, the right plot shows that less efficient solvers like $\mathcal S_2$ together with appropriate short updating periods can satisfy the decreasing condition (\ref{ineqtradeoff}). The right figure also shows that in this latter case, there may be several possible values of $\tau_u$ (several number of iterations) that may decrease the cost and an adaptive on-line monitoring algorithm like the one recalled in section \ref{recalalgo} may be appropriate to get closer to an optimal decrease. \ \\ \ \\ 
In the following sections, the two solvers that are used in the validation section are introduced. 
\vskip 0.5cm 
\section{Presentation of the algorithms} \label{secalgo} 
\subsection{qpOASES}\label{secalgoasis} 

The qpOASES \cite{Ferreau2014} solver is a well know solver in the linear constrained MPC control community. It offers a very efficient implementation of the active-set strategy \cite{Ferreau2008}. If several QP problems must be solved with constant Hessian and constraint matrices, the qpOASES package offers the possibility of hot-starting from previous solution with a subroutine called \textit{qpOASES\textunderscore sequence}. In the sequel, the \textit{qpOASES\textunderscore sequence} subroutine will be used and will simply be recalled as qpOASES.

\subsection{ODE-based solver}\label{odebasedsolver}
In this section, an ODE-based solver that is used hereafter to implement the PLC-based constrained MPC is briefly presented. The real-time performance of this solver is also compared to that of qpOASES in the PLC constrained performance setting in order to illustrate Fact 1 mentioned above. \ \\ \ \\ 
Consider the Quadratic Programming (QP) problem defined by:
\begin{equation}\label{pb_normal}
\tilde{\mathcal{P}}(z) = \left\lbrace \matl{ \text{min:~} J_0(z) = z^T \Phi z + z^T \phi \cr \text{under~} \left\lbrace \matl{\Gamma z - \gamma \le 0 \cr \underline{z} \le z \le \overline{z} } \right. } \right. 
\end{equation}
where $z \in \mathbb{R}^{n_z}$ is the decision variable while $\Phi$ and $\phi$ are matrices of appropriate size. $\Gamma\in \mathbb{R}^{n_c\times n_z} $ and $\gamma\in \mathbb{R}^{n_c}$ are the matrices that define the set of $n_c$ inequality constraints while $\underline{z}$ and $\overline{z}$ are lower and upper bounds on the decision variables. \\ \ \\ 
Based on the above formulation, the following augmented cost function can be defined:
\begin{eqnarray}
J(z):=J_0(z)&&+\alpha\sum_{i=1}^{n_c}\max(\Gamma_iz-\gamma_i,0)^\mu+\nonumber \\
&&+\alpha\sum_{i=1}^{n_z}\max(z_i-\overline{z}_i,0)^\mu+\nonumber \\
&&+\alpha\sum_{i=1}^{n_z}\max(\underline{z}_i-z_i,0)^\mu \label{augmentedJ} 
\end{eqnarray} 
where $\Gamma_i\in \mathbb{R}^{1\times n_z}$ is the $i$-th line of $\Gamma$. 
Based on this augmented cost, the following Ordinary Differential Equation (ODE) can be used to define a trajectory in the decision variable space along which the augmented cost decreases:
\begin{eqnarray}
\dot z=-\dfrac{dJ}{dz}(z) \label{TheOde} 
\end{eqnarray} 
Note however that this ODE is generaly stiff because of the high values of $\alpha$ one needs to use in order to enforce the constraints fulfillment. That is the reason why the one-step Backward-Differentiation-Formulae (TR-BDF2) described in \cite{trbdf2} for stiff differential equations is used here. \ \\ \ \\ 
Note also that after the computation of the TR-BDF2 step, all the decision variable that correspond to hard constraints (saturation on actuator for instance) are projected into the admissible box before a next iteration is computed. In addition to the integration scheme described in \cite{trbdf2}, the initial time step is defined by using the following expression:
\begin{equation}
\Delta t = \sqrt{\dfrac{1}{\| \dot{z}(t) \|}}
\end{equation}

In the case of the quadratic problem described in paragraph \ref{Theproblem}, this method leads to fast convergence to the suboptimal solution $z^*$, being very close to the actual optimal solution of the original problem even with real-time constraints. The comparison between solvers \ref{secalgoasis} and \ref{odebasedsolver} will be done in paragraph \ref{secfastexpcomp}.\ \\ \ \\ 
Note also that this solver fully satisfies the decrease condition (\ref{decrease}) since it moves along the descent trajectory defined by (\ref{TheOde}). Therefore, the adaptation mechanism of the control updating period can be applied. 
\section{Plant description} \label{secplant} 
\subsection{General presentation}
\begin{figure*}
\begin{center}
\includegraphics[width=0.90\textwidth]{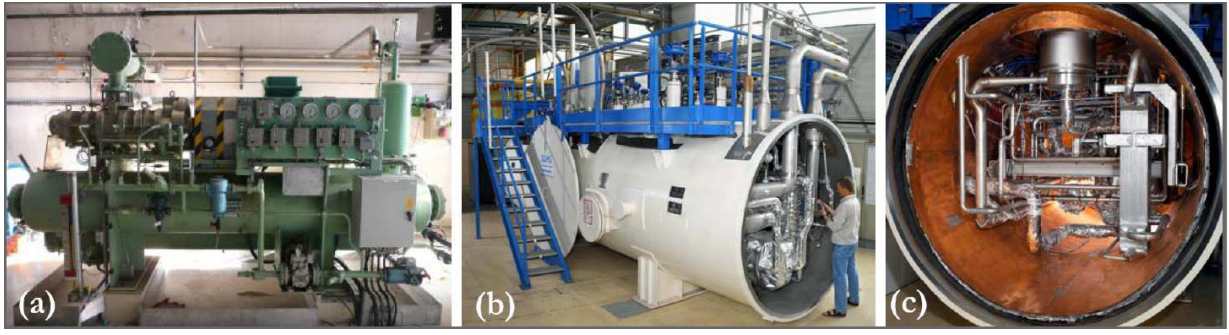}
\caption{Views of the cryogenic plant of CEA-INAC-SBT, Grenoble. (a) The screw compressor of the warm compression station. (b) The cold box. (c) Internal detail of the cold box.}\label{400W_photos}
\end{center}
\end{figure*}

\begin{figure}
\begin{center}
\begin{psfrags}

\psfrag{name114}{\small{$\mathcal{S}1$}}
\psfrag{name111}{\small{$\mathcal{S}2$}}
\psfrag{name112}{\small{$\mathcal{S}3$}}
\psfrag{name113}{\small{$\mathcal{S}4$}}

\psfrag{name101}{\small{$NS_1$}}
\psfrag{name102}{\small{$CV_{155}$}}
\psfrag{name103}{\small{$NEF_1$}}
\psfrag{name104}{\small{$NEF_2$}}
\psfrag{name105}{\small{$Stt207$}}
\psfrag{name106}{\small{$NEF_{34}$}}
\psfrag{name107}{\small{$CV_{156}$}}
\psfrag{name108}{\small{$NEF_5$}}
\psfrag{name109}{\small{$NEF_6$}}
\psfrag{name110}{\small{$CV_{167}$}}

\psfrag{tag113}{\small{$LN2$}}
\psfrag{tag114}{\small{$GN2$}}

\psfrag{tag119}{\small{$CV_{952}$}}
\psfrag{tag120}{\small{$CV_{953}$}}
\psfrag{tag121}{\small{$CV_{956}$}}
\psfrag{tag122}{\small{$NC_1+NC_2$}}

\psfrag{tag105}{\small{$T_3,~P_3$}}

\psfrag{tag107}{\small{$T_4,~P_4$}}
\psfrag{tag108}{\small{$T_5,~P_5$}}

\psfrag{tag109}{\small{$T_6,~P_6$}}
\psfrag{tag110}{\small{$T_7,~P_7$}}
\psfrag{tag106}{\small{$T_8,~P_8$}}

\psfrag{tag111}{\small{$T_9,~P_9$}}
\psfrag{tag112}{\small{$T_{10},~P_{10}$}}

\psfrag{tag115}{\small{$T_{11},~P_{11}$}}
\psfrag{tag116}{\hspace{-0.8cm}\small{$T_{12},~P_{12},~M_{12}$}}
%\psfrag{tag117}{\small{$ $}}
%\psfrag{tag118}{\small{$ $}}
\psfrag{tag117}{\small{$S_1$}}

\psfrag{tag101}{\small{$NCR_{22}$}}
\psfrag{tag102}{\small{$L_1,~T_1,~P_1$}}
\psfrag{tag103}{\small{$ $}}
\psfrag{tag104}{\small{$T_2,~P_2$}}

\includegraphics[scale = .68]{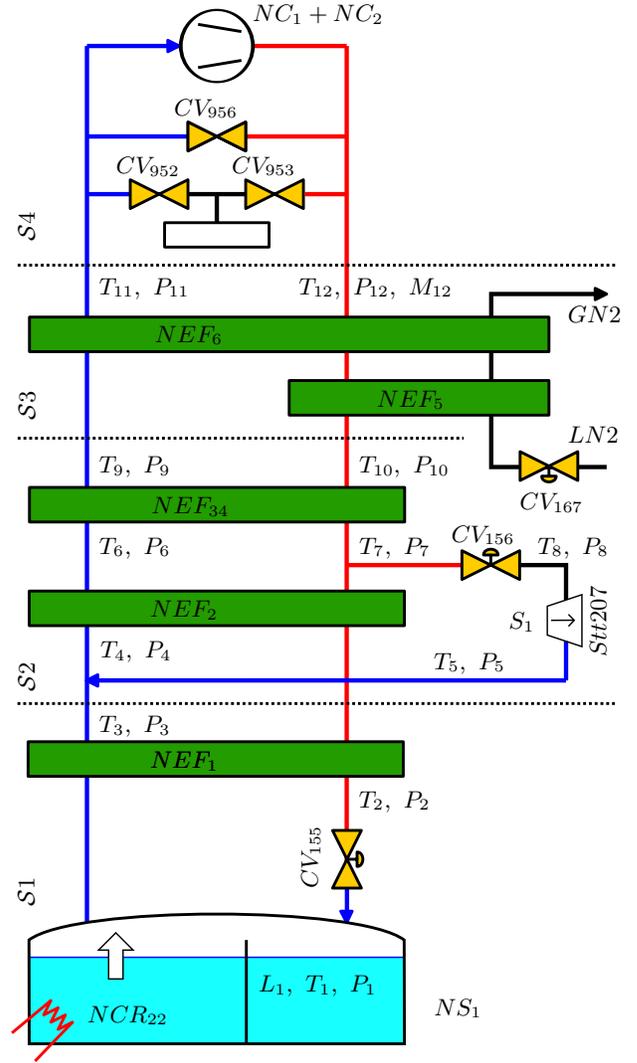}

\end{psfrags}
\caption{Functional overview of the $450 W$ at $4.4K$ helium refrigerator available at CEA-INAC-SBT, Grenoble. The components named $CV$ are controlled valves, used to control the system. The label $Stt$ stands for the cryogenic turbine while $NS$ is used for the phase separator. $NC$'s are helium compressors while $NEF$'s stand for heat exchangers. $T$'s and $P$'s stand for temperature and pressure sensors. $S_1$ is the turbine speed sensor while $L_1$ stands for the bath level sensor.}
\label{400W_schema}
\end{center}
\end{figure}

Fig. \ref{400W_photos} shows an overview of the cryogenic plant of the CEA-INAC-SBT, Grenoble. This plant provides a nominal cooling capacity of $450\ W$  at $4.4\ K$ in the configuration in which this study have been done. It is dedicated to physical experiments (cryogenic component testing, turbulence and pulsed heat load studies, etc.).\\

The process flow diagram of the cryogenic plant is shown in Fig. \ref{400W_schema}. One may notice the following main elements:
\begin{itemize}
\item[-] Two volumetric screw compressors ($NC_*$) and a set of control valves ($CV_{95*}$),
\item[-] Several counterflow heat exchangers ($NEF_*$), a liquid nitrogen pre-cooler ($NEF_5$), 
\item[-] A cold turbine expander which extracts work from the circulating gas ($Stt_{207}$),
\item[-] A so-called turbine valve ($CV_{156}$),
\item[-] A Joule-Thomson expansion valve for helium liquefaction ($CV_{155}$),
\item[-] A phase separator ($NS_1$), connected to the loads (simulated here by the heating device referred as $NCR_{22}$).\\
\end{itemize}
Note that the plant can be viewed as the interconnection of four elementary subsystems: the Warm Compression Station ($\mathcal{S}_4$), the Nitrogen Pre-Cooler ($\mathcal{S}_3$), the Brayton Cycle ($\mathcal{S}_2$) and the JT cycle ($\mathcal{S}_1$), delimited by dotted lines in Fig.  \ref{400W_schema}. While constrained MPC is used in this study, the cryogenic system is classically controlled by three independent control loops:
\begin{itemize}
\item The output temperature of the turbine expander is controlled with a PI controller working with the turbine valve $CV_{156}$;
\item the level of liquid helium in the tank is controller by a PI controller, working with the heating device $NCR_{22}$;
\item the high and low pressures (in red and blue pipes, respectively) is controlled by an LQ controller, like the one described in \cite{bonne_cec_1};
\end{itemize} 
the valve $CV_{155}$ being used at a constant opening set by the user, depending on the application. In this study, attention has been focused on subsystems $1$ and $2$, with are the coldest part of the refrigerator (from $80K$ to $4.4K$). More informations about the plant can be found in \cite{clavel_these}.

\subsection{Model derivation and properties}
In order to derive the system model, several studies have been conducted  \cite{clavel_these,clavel,bonne:hal-00922066,bonne_cec_2}. The \JT cycle of this paper has been modelled in \cite{bonne:hal-00922066} while the Brayton cycle is presented in details in \cite{bonne_cec_2}. It is worth mentioning that heat exchangers involve models with coupled partial differential equations (PDEs) that have been spatially discretized, leading to rather large state space. In this study, the two models has been merged to obtain a state space model that takes the following form:
\begin{subequations}\label{model_cold_end}
\begin{align}
\dot{\bx} &= f^1(\bx,\bu,\bw)\\
\by &= f^2(\bx,\bu,\bw)
\end{align}
\end{subequations}
where $f^1$ is the function that express the derivative of the state $\bx$  while $f^2$ is the function that express the measured output vector $\by$. Both functions are continuously differentiable. State vector, input vector, and disturbance vector are expressed more precisely by
\begin{equation}
\bx = \pmat{\bx_{ns1}\cr \bx_{nef2}\cr \bx_{nef1}}, \quad \bu = \pmat{CV_{155}  \cr NCR_{22}^{A} \cr CV_{156}}, \quad \bw = NCR_{22}^{HL}
\end{equation}
where $\bx_{ns1}$, $\bx_{nef1}$ and $\bx_{nef2}$ depict the state vector of individual components, described in \cite{bonne:hal-00922066,bonne_cec_2}. It has to be noted that $NCR_{22}$ is used both to control the plant and to disturb it. That is why it has been named $NCR_{22}^{A}$ for the actuator part and $NCR_{22}^{HL}$ for the heat load part. At the end, $NCR_{22} = NCR_{22}^{HL} + NCR_{22}^{A}$. The vector of measured output is the following:
\begin{equation}
\by = \pmat{L_1 & V_1 & T_1 & \cdots & T_{10} & P_1 & \cdots & P_{10}}^T
\end{equation}
It has been shown in \cite{bonne_cec_2} that the non-linear model expressed by \eqref{model_cold_end} can be linearized around an operation point of interest defined by  $f^1(\bx_0,\bu_0,\bw_0) = 0$. The linearized model is then discretized using Matlab function $c2d(\cdot)$ with sampling period $\tau = 5s$, leading to the following discrete LTI model:
\begin{eqnarray}
\tilde{\bx}_{k+1} = A \tilde{\bx}_{k} + B \tilde{\bu}_{k} + F \tilde{\bw}_{k} \\\label{gftr56} 
\tilde{\by} = C \tilde{\bx}_{k} + D \tilde{\bu}_{k} + G \tilde{\bw}_{k}\label{gftr57} 
\end{eqnarray}
where variables with a tilde depict the deviation of the original variables around the operating point of interest:
\begin{equation}\label{model_cold_end}
\begin{split}
\bx_{k} & = \tilde{\bx}_{k} +\bx_0, \quad \tilde{\bu}_{k} = \bu_{k} - \bu_0\\
\by_{k} & = \tilde{\by}_{k} +\by_0, \quad \tilde{\bw}_{k} = \bw_{k} - \bw_0
\end{split}
\end{equation}
Note that the model defined by (\ref{gftr56}) stands for the prediction model (\ref{predictionmodel}) invoked in the general presentation of MPC (section \ref{defandnotes}). Following the same notation, the predicted output is denoted by $\by_{k+j}=Y(j,\bx_k,p,\tilde\tbw_k)$ while the true measured output is denoted by $Y^r(j,\bx_k,p,\tilde\tbw_k)$. 

\subsection{Statement of the MPC-related optimisation problem}\label{Theproblem}
\begin{figure}[H]
\input{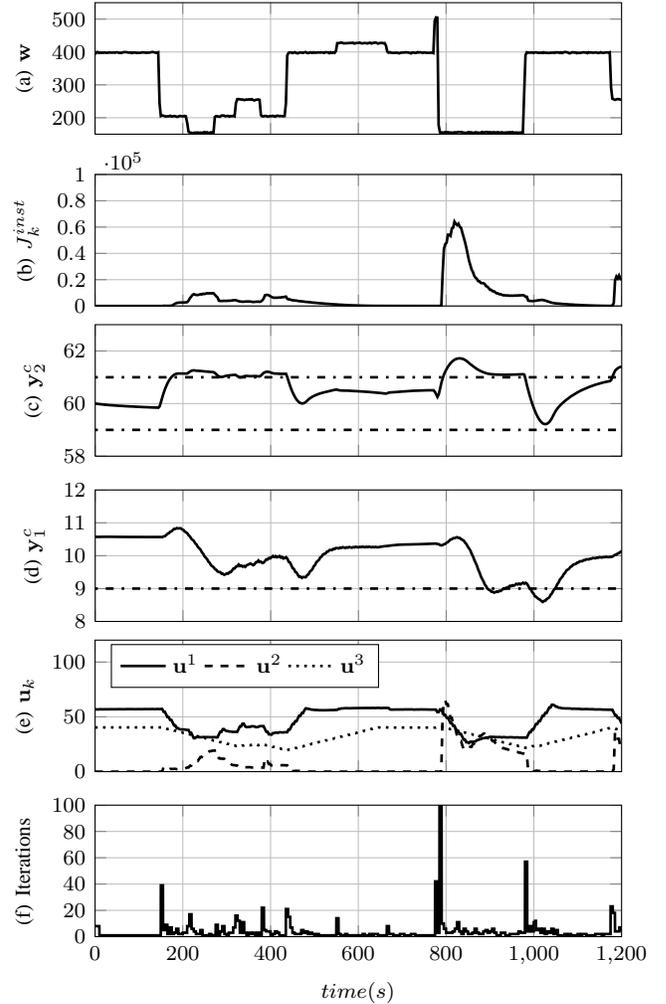}
\caption{Simulated behaviour of the system under qpOASES-based MPC control without limitation of the number of iterations.}
\label{working_mpc}
\end{figure}

\noindent First of all, the following constraints have to be satisfied as far as possible:
\begin{subequations}\label{contr}
\begin{align}
\underline{y^c} \leqslant ~ &\by^c_k \leqslant \overline{y^{c}}\\
\underline{\bu} \leqslant ~ &\bu_k \leqslant \overline{\bu}\\
\underline{\delta \bu } \leqslant  ~ &\delta \bu_k \leqslant \overline{\delta \bu } 
\end{align}
\end{subequations}
where $\delta \bu_k$ stand for the increment $\bu_k-\bu_{k-1}$ on the input vector. $\by^c_k $ denotes a subset of output components $\by_k $ which is constrained. This subset is composed of the helium bath level $L_1$ and the turbine output temperature $T_5$. Details regarding the variables involved in (\ref{contr}) are given in table \ref{tabletable}:
\renewcommand{\arraystretch}{1.2}
\begin{table}[H]
\begin{center}
\begin{tabular}{c|c|c}
Var. & Meaning & Value \\
\hline
$\underline{\bu}$ & min. control effort & $ ( 20~~20~~0 )^T $ \\
$\overline{\bu}$ & max. control effort & $ ( 60~~60~~150 )^T $  \\
$\underline{y^c}$ & low limit on the output & $ ( 59~~16 )^T $ \\
$\overline{y^c}$ & high limit on the output & $ ( 61~~9 )^T $ \\
$\underline{\delta \bu }$ & max increment & $ ( 0.5~~10~~0.1 )^T $ \\
$\overline{\delta \bu }$ & min increment & $ ( 0.5~~10~~0.1 )^T $ \\
\end{tabular}
\end{center}
\caption{The constraints bounds} \label{tabletable} 
\end{table}
One of the specific feature of Output constraints is that they cannot be necessarily fully respected depending on the unpredictable thermal loads. That is why these constrained are systematically relaxed.  This is introduced through the constraint violation variable $\bv_k$ that is defined as follows:
\begin{equation}\label{inst_viol}
\bv_k = max(\by^c_k-\overline{y^{c}},0) +  max(\underline{y^c} - \by^c_k,0)
\end{equation}
while constraint violation prediction at sampling instant $k+j$ is written:
\begin{equation}\label{traj_viol}
\begin{split}
V(j,\bx_k,p,\hat\tbw_k) = max(Y^c(j,\bx_k,p,\hat\tbw_k)-\overline{y^{c}},0) + \\ max(\underline{y^c} -Y^c(j,\bx_k,p,\hat\tbw_k),0)
\end{split}
\end{equation}
where $Y^c(j,\bx_k,p,\mathcal{W})$ is used to define the constrained subset of $Y(j,\bx_k,p,\mathcal{W})$.\ \\ \ \\ 
The sequence of control vectors $u^{(i)} (p)$ is then obtained by minimizing the cost function :
\begin{equation}\label{real_cost}
\begin{split}
J_k = \sum _{j=1}^{N_p}  \|X(j,\bx_k,p,\hat\tbw_k)\|_Q^2 + \|u^{(j)}(p)\|_R^2 + \\  \|V(j,\bx_k,p,\hat\tbw_k))\|_\rho^2 
\end{split}
\end{equation}
where $Q$ and $R$ are weighting matrices on the state and input vectors while $\rho$ defines the constraint violation-related penalty. This cost function, together with the linear constrained and the linearized model (\ref{gftr57}) lead to a constrained QP problem which if of the form (\ref{pb_normal}) in which the decision variable $z$ is precisely the control profile parameter $p$.  Note also that the affine term $\phi$ [see (\ref{pb_normal})] does depend on the current value of the disturbance $\bw=NCR_{22}^{HL}$. 
\ \\ \ \\ 
By choosing a sampling period $\tau = 5$ sec, preliminary simulations showed that a prediction horizon of at least $N_p = 100$ is required. This leads to an optimization problem that involves $700$ decision variables and a total number of $1000$ constraints to be satisfied if trivial pwc parametrization is adopted. Such problem are beyond the computational capacity of the targeted industrial PLC (see the performance of our PLC in the section \ref{PLCDES}).\ \\ \ \\ 
To reduce the problem dimension, the control profile has been parametrized using classic piece-wise affine method that leaves as decision variables the values of the control inputs at $7$ decisions instants\footnote{decisions instants are chosen to be: $(1,~2,~4,~8,~16,~50,~50,~100)$}. Moreover, the constraints satisfaction is checked only at $14$ future instants\footnote{constraints verifications instants are chosen to be:\\  $(1,~2,~3,~4,~6,~8,~16,~24,~32,~48,~60,~72,~84,~100)$}. This finally leads to an optimization problem involving $49$ decision variables (note that there are $7$ control inputs, namely $3$ physical input and $4$ virtual input representing the constraints violation), with $56$ (outputs) plus $38$ rate saturation constraints to be satisfied. \ \\ \ \\ 
To ensure that this scheme is appropriate to control the plant, the problem closed-loop system is first simulated using the qpOASES solver. Time results are presented in Fig. \ref{working_mpc}. Fig. \ref{working_mpc} (a) shows the thermal heat load that has been used in this simulation. Part (b) shows that the scheme is able to decrease the stage instantaneous cost define as:
\begin{equation}\label{cost_dev_bo}
\bar{J}^{inst}_k= \|\bx_k^r\big\|^2_Q + \|\bu_k\|^2_R  +\|\bv_k\big\|^2_\rho
\end{equation}
Parts (c) and (d) of Fig. \ref{working_mpc} show that the constraints are violated within limited amplitude and duration. Part (e) shows the control effort. Part (f) shows the number of iterations of the qpOASES solver. It is worth mentioning that the number of iterations is important during heat load event. This has significant consequence on real-time feasibility of the qpOASES-based solution  as it is examined in the sequel.

\setlength\fheight{1.75cm}
\setlength\fwidth{7cm}

\subsection{Description of the PLC} \label{PLCDES} 
\begin{figure}[H]
\begin{center}
\includegraphics[width=0.1\textwidth]{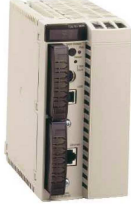}
\caption{Schneider PLC TSX P574634M}
\label{tsx}
\end{center}
\end{figure}

\begin{figure}[t]
\input{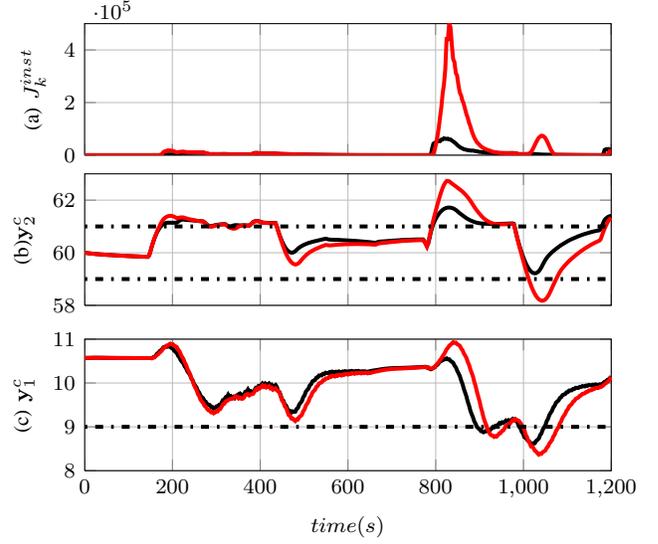}
\caption{Simulated behaviour of the system under MPC control for both unconstrained (black lines) and constrained (red lines) solving time}
\label{not_working_mpc}
\end{figure}

\noindent This section focuses on the Programmable Logic Controller (PLC) available to implement the QP-based constrained MPC. It is a Schneider TSX P574634M shown in Fig. \ref{tsx}. This PLC is fully dedicated for our application and it communicates optimisation results to another PLC that actually controls the plant.
\ \\ \ \\ 
According to the manufacturer, this PLC shows maximum computing capability of about $1.8~Mflops$ \cite{doc_plc}. In order to evaluate this claim, the Cholesky factorisation of increasing size matrices has been executed while monitored the computation times. Fig. \ref{performance_automate} shows the results compares them to the performance of a nowadays DELL Latitude E6520 laptop with Intel I5-2520M CPU. This graph shows a slowing factor that lies around $4000$. Note that the same graph shows the performance of the PLC in ms while the performance of the desk computer is shown in $\mu$s.\ \\ \ \\ 
Note that the PLC is used with an external PCMCIA memory card of $2Mb$, shared for both code and variables. This makes memory also a crucial issue. Indeed without reduced parametrization, the Hessian of the QP problem would have just fit the memory size of the PLC, since it represents a total memory occupation $4*700^2 = 1.96Mb$ in single precision representation. 
\setlength\fheight{2.5cm}
\setlength\fwidth{7cm}
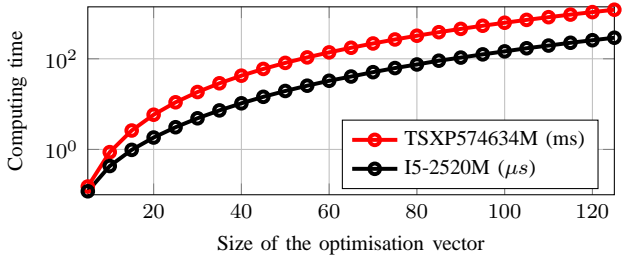
\begin{figure}[t]
% This file was created by matlab2tikz v0.4.7 running on MATLAB 7.14.
% Copyright (c) 2008--2014, Nico SchlÃ¶mer <nico.schloemer@gmail.com>
% All rights reserved.
% Minimal pgfplots version: 1.3
% 
% The latest updates can be retrieved from
%   http://www.mathworks.com/matlabcentral/fileexchange/22022-matlab2tikz
% where you can also make suggestions and rate matlab2tikz.
% 
\begin{tikzpicture}

\begin{axis}[%
width=\fwidth,
height=\fheight,
scale only axis,
xmin=5,
xmax=125,
xlabel={Size of the optimisation vector},
xmajorgrids,
ymode=log,
ymin=0.1,
ymax=1400,
yminorticks=true,
ylabel={Computing time},
ymajorgrids,
yminorgrids,
legend style={at={(0.97,0.03)},anchor=south east,draw=black,fill=white,legend cell align=left}
]
\addplot [color=red,solid,line width=1.5pt,mark=o,mark options={solid}]
  table[row sep=crcr]{5	0.151888888888889\\
10	0.871764705882353\\
15	2.614\\
20	5.828\\
25	10.968\\
30	18.4666666666667\\
35	28.8\\
40	42.3823529411765\\
45	59.7083333333333\\
50	81.1666666666667\\
55	107.307692307692\\
60	138.4\\
65	175.125\\
70	217.666666666667\\
75	266.8\\
80	322.75\\
85	386\\
90	457\\
95	536\\
100	624\\
105	721.5\\
110	828\\
115	944\\
120	1071\\
125	1209\\
};
\addlegendentry{TSXP574634M (ms)};

\addplot [color=black,solid,line width=1.5pt,mark=o,mark options={solid}]
  table[row sep=crcr]{5	0.1192400011924\\
10	0.430320017212801\\
15	0.977242619712221\\
20	1.83008029281285\\
25	3.08906153466827\\
30	4.86576311408839\\
35	7.26376116220179\\
40	10.4083266613291\\
45	14.4781442735727\\
50	19.485\\
55	25.4799324092543\\
60	32.7411830860981\\
65	40.5900473242831\\
70	50.6385681021709\\
75	62.4235350367735\\
80	74.8635516506922\\
85	90.7299754203894\\
90	106.277782817745\\
95	125.130862646566\\
100	145.681456814568\\
105	170.750880071089\\
110	195.81042360412\\
115	225.592458429712\\
120	256.842105263158\\
125	291.315760691239\\
};
\addlegendentry{I5-2520M ($\mu s$)};

\end{axis}
\end{tikzpicture}%
\caption{Cholesky factorisation time for two different CPUs. It can be noticed that the performance ratio between the PLC and the laptop is about $4000$ for matrices sized $40$ to $125$.}
\label{performance_automate}
\end{figure}
\ \\ \ \\ Now since a single iteration of the qpOASES solver takes approximately $120\mu s$, the same iteration would take $0.12*4000 = 0.48 s$ when executed on the PLC. Therefore only $10$ iterations of the qpOASES solver can be performed during the sampling period $\tau=5$ sec. The scenario that has been shown in Fig. \ref{working_mpc} with no bound on the number of iterations has been simulated with the qpOASES 'maxiter' option set to $10$. The result is presented by Fig. \ref{not_working_mpc} on which the unlimited case has been also reported for easiness of comparison. 
\setlength\fheight{1.75cm}
\setlength\fwidth{7cm}
\ \\ \ \\ 
Figure \ref{not_working_mpc} shows that when the number of iterations of the hot-started qpOASES solver is limited to $10$, the closed-loop performance as well as the constraints fulfillment are drastically affected. This is precisely for this reason that the ODE-based solver explained in section \ref{odebasedsolver} has been developed since it corresponds to a less computation time per iteration and can  therefore be potentially more suitable in presence of the limited performance available PLC following the discussion of section \ref{fundtradeoff}. 
\section{Fast MPC-related investigation} \label{secfastexp} 
\subsection{Comparison of algorithms} \label{secfastexpcomp} 
\begin{figure}
\input{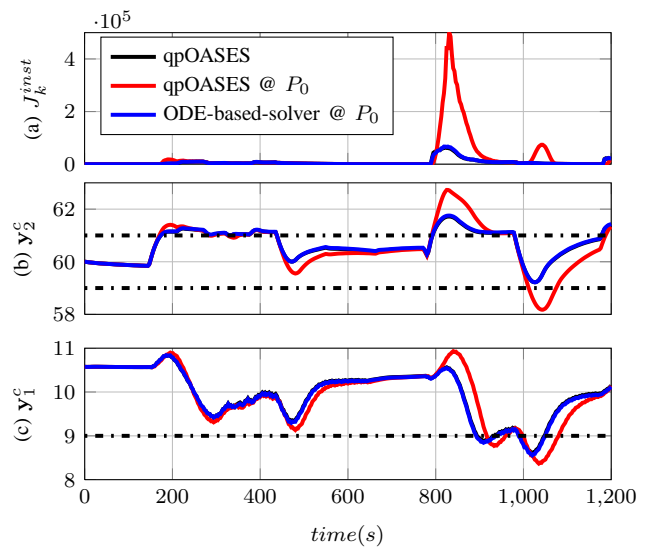}
\caption{Comparison of the closed-loop performance under the ODE-based solver ($20$ iterations), the qpOASES solver ($10$ iterations) and the qpOASES (without limitations).}
\label{mult_solv}
\end{figure}

\noindent The aim of the present section is to assess the first Fact mentioned above, namely that it is sometimes better to use a less efficient per iteration solver (the ODE-based solver in our case) provided that it corresponds to less computation time per iteration. In our case, as far as the above described PLC is used, it is possible to perform $20$ iterations of the ODE-based solver against only $10$ iterations of the qpOASES solver. \ \\ \ \\  
Eight hours simulations have been done with the two solvers, with a variable computational capability (i.e. a variable allowed number of iterations). Some relevant results are plotted, always as a function of the normalized computation capability $\bar{P} = P/P_0$ where $P_0$ is the computational capability of our device.\ \\ \ \\ 
In order to support the comparison that can be difficult because of the presence of relaxed weighted constraints, the cost (\ref{real_cost}) to be minimized at each sampling period has been divided in two separated parts, in order to compare them separately. The first part represents the deviation cost:
\begin{equation}\label{cost_dev_bo}
\bar{J}^{dev}_k = \sum _{j=1}^{N_p}  \|X(j,\bx_k,p,\hat\tbw_k)\|_Q^2 +  \|u^{(j)}(p)\|_R^2 
\end{equation}
while the second part stands for the outputs constraints violation cost:
\begin{equation}\label{cost_cst_bo}
\bar{J}^{cst}_k = \sum _{j=1}^{N_p}  \|V(j,\bx_k,p,\hat\tbw_k)\|_\rho^2
\end{equation}
and then the sum of those two costs along the whole simulation is expressed:\\
\begin{minipage}[r]{0.243\textwidth}
\begin{equation}\label{cost_sum_bo_dev} 
\bar{\mathcal{J}}^{dev} = \sum _{k=1} ^{N_{sim}} \bar{J}_k^{dev} 
\end{equation}
\end{minipage}
\hfill
\begin{minipage}[l]{0.23\textwidth}
\begin{equation}\label{cost_sum_bo_cst} 
\bar{\mathcal{J}}^{cst} = \sum _{k=1} ^{N_{sim}} \bar{J}_k^{cst} 
\end{equation}
\end{minipage}
~\\ where $N_{sim}$ is the number of problems solved during the simulation.\\
\setlength\fheight{1.75cm}
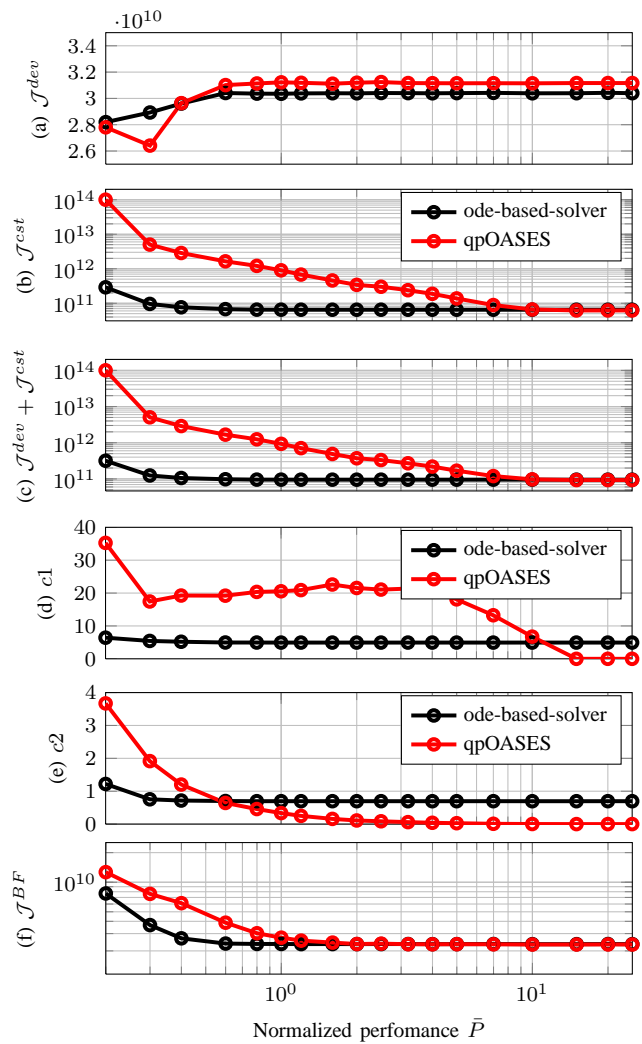
\begin{figure}
% This file was created by matlab2tikz v0.4.7 running on MATLAB 7.14.
% Copyright (c) 2008--2014, Nico SchlÃ¶mer <nico.schloemer@gmail.com>
% All rights reserved.
% Minimal pgfplots version: 1.3
% 
% The latest updates can be retrieved from
%   http://www.mathworks.com/matlabcentral/fileexchange/22022-matlab2tikz
% where you can also make suggestions and rate matlab2tikz.
% 
\begin{tikzpicture}

\begin{axis}[%
width=\fwidth,
height=\fheight,
scale only axis,
xmode=log,
xmin=0.2,
xmax=25,
xtick={1,10},
xticklabels={\empty},
xminorticks=true,
xmajorgrids,
xminorgrids,
ymode=log,
ymin=30949417872.2853,
ymax=200745086258463,
ytick={     1000000000,     10000000000,    100000000000,   1000000000000,  10000000000000, 100000000000000},
yminorticks=true,
ylabel={ (b) $\mathcal{J}^{cst}$},
ymajorgrids,
yminorgrids,
name=plot2,
legend style={draw=black,fill=white,legend cell align=left}
]
\addplot [color=black,solid,line width=1.5pt,mark=o,mark options={solid}]
  table[row sep=crcr]{0.2	289949321100.102\\
0.3	95708072783.6066\\
0.4	76424830914.9735\\
0.6	67572501757.536\\
0.8	65377915523.5399\\
1	65166344692.0665\\
1.2	65041214795.834\\
1.6	64998161989.0669\\
2	64986019237.5488\\
2.5	65025912431.351\\
3.2	65026487435.8295\\
4	65013012021.0389\\
5	65030132759.779\\
7	65033818440.8237\\
10	65019679645.5328\\
15	65019241412.7442\\
20	65036885178.7517\\
25	65023710088.7788\\
};
\addlegendentry{ode-based-solver};

\addplot [color=red,solid,line width=1.5pt,mark=o,mark options={solid}]
  table[row sep=crcr]{0.2	100372543129231\\
0.3	5015999868385.41\\
0.4	2861147915468.78\\
0.6	1640801085397.59\\
0.8	1212350931260.47\\
1	901357038857.83\\
1.2	678103943025.73\\
1.6	458301283801.099\\
2	341186064831.376\\
2.5	301215128706.339\\
3.2	239181533250.502\\
4	187880793402.432\\
5	137813232709.466\\
7	88734141131.7824\\
10	66078392005.6547\\
15	61898835744.5706\\
20	61898835744.5706\\
25	61898835744.5706\\
};
\addlegendentry{qpOASES};

\end{axis}

\begin{axis}[%
width=\fwidth,
height=\fheight,
scale only axis,
xmode=log,
xmin=0.2,
xmax=25,
xtick={1,10},
xticklabels={\empty},
xminorticks=true,
xmajorgrids,
xminorgrids,
ymin=25000000000,
ymax=35000000000,
ylabel={ (a) $\mathcal{J}^{dev}$},
ymajorgrids,
at=(plot2.above north west),
anchor=below south west
]
\addplot [color=black,solid,line width=1.5pt,mark=o,mark options={solid},forget plot]
  table[row sep=crcr]{0.2	28195522874.4284\\
0.3	28928677573.3971\\
0.4	29607953026.0623\\
0.6	30410124492.647\\
0.8	30363284145.4878\\
1	30353487343.9428\\
1.2	30381361726.2201\\
1.6	30391936879.992\\
2	30385602178.7552\\
2.5	30410102669.3308\\
3.2	30402152331.0041\\
4	30389096038.2533\\
5	30402487179.8384\\
7	30421193910.0616\\
10	30388442933.8386\\
15	30394232257.3429\\
20	30425110084.599\\
25	30397322033.1784\\
};
\addplot [color=red,solid,line width=1.5pt,mark=o,mark options={solid},forget plot]
  table[row sep=crcr]{0.2	27802236662.6772\\
0.3	26413088385.554\\
0.4	29639035869.2045\\
0.6	31013130325.3099\\
0.8	31136684220.3581\\
1	31221620428.4777\\
1.2	31187329289.5311\\
1.6	31118757837.9304\\
2	31193067678.0879\\
2.5	31241321651.8781\\
3.2	31166737374.3809\\
4	31156359078.3653\\
5	31151499840.6208\\
7	31155730510.7787\\
10	31147811305.3377\\
15	31167664577.5831\\
20	31167664577.5831\\
25	31167664577.5831\\
};
\end{axis}

\begin{axis}[%
width=\fwidth,
height=\fheight,
scale only axis,
xmode=log,
xmin=0.2,
xmax=25,
xtick={1,10},
xticklabels={\empty},
xminorticks=true,
xmajorgrids,
xminorgrids,
ymode=log,
ymin=46533250161.0768,
ymax=200800690731788,
ytick={     1000000000,     10000000000,    100000000000,   1000000000000,  10000000000000, 100000000000000},
yminorticks=true,
ylabel={(c) $\mathcal{J}^{dev}+\mathcal{J}^{cst}$},
ymajorgrids,
yminorgrids,
name=plot3,
at=(plot2.below south west),
anchor=above north west
]
\addplot [color=black,solid,line width=1.5pt,mark=o,mark options={solid},forget plot]
  table[row sep=crcr]{0.2	318144843974.53\\
0.3	124636750357.004\\
0.4	106032783941.036\\
0.6	97982626250.1831\\
0.8	95741199669.0278\\
1	95519832036.0093\\
1.2	95422576522.0542\\
1.6	95390098869.0589\\
2	95371621416.3039\\
2.5	95436015100.6818\\
3.2	95428639766.8336\\
4	95402108059.2922\\
5	95432619939.6175\\
7	95455012350.8854\\
10	95408122579.3714\\
15	95413473670.0871\\
20	95461995263.3506\\
25	95421032121.9572\\
};
\addplot [color=red,solid,line width=1.5pt,mark=o,mark options={solid},forget plot]
  table[row sep=crcr]{0.2	100400345365894\\
0.3	5042412956770.97\\
0.4	2890786951337.98\\
0.6	1671814215722.9\\
0.8	1243487615480.83\\
1	932578659286.308\\
1.2	709291272315.261\\
1.6	489420041639.029\\
2	372379132509.464\\
2.5	332456450358.217\\
3.2	270348270624.883\\
4	219037152480.798\\
5	168964732550.087\\
7	119889871642.561\\
10	97226203310.9924\\
15	93066500322.1536\\
20	93066500322.1536\\
25	93066500322.1536\\
};
\end{axis}

\begin{axis}[%
width=\fwidth,
height=\fheight,
scale only axis,
xmode=log,
xmin=0.2,
xmax=25,
xtick={1,10},
xticklabels={\empty},
xminorticks=true,
xmajorgrids,
xminorgrids,
ymin=0,
ymax=40,
ylabel={ (d) $c1$},
ymajorgrids,
name=plot4,
at=(plot3.below south west),
anchor=above north west,
legend style={draw=black,fill=white,legend cell align=left}
]
\addplot [color=black,solid,line width=1.5pt,mark=o,mark options={solid}]
  table[row sep=crcr]{0.2	6.42795733874094\\
0.3	5.44239965130666\\
0.4	5.19480203214249\\
0.6	4.96391088133887\\
0.8	4.95348890162409\\
1	4.93905471612145\\
1.2	4.94274497551324\\
1.6	4.94037595330754\\
2	4.94335015625682\\
2.5	4.94153691468957\\
3.2	4.9420052712448\\
4	4.94375418431343\\
5	4.94250939325354\\
7	4.93812287724362\\
10	4.94393880795295\\
15	4.94378461119062\\
20	4.93878287192706\\
25	4.94332564716911\\
};
\addlegendentry{ode-based-solver};

\addplot [color=red,solid,line width=1.5pt,mark=o,mark options={solid}]
  table[row sep=crcr]{0.2	35.2585732900486\\
0.3	17.4379233680065\\
0.4	19.2477666355179\\
0.6	19.2091490111904\\
0.8	20.3479151796257\\
1	20.5415030299906\\
1.2	20.8993054106453\\
1.6	22.5990278085348\\
2	21.5447741527484\\
2.5	21.0596190088353\\
3.2	21.348357113695\\
4	21.172165352554\\
5	18.0857586482315\\
7	13.2498375380241\\
10	6.74187911830572\\
15	1.79856129989275e-14\\
20	1.79856129989275e-14\\
25	1.79856129989275e-14\\
};
\addlegendentry{qpOASES};

\end{axis}

\begin{axis}[%
width=\fwidth,
height=\fheight,
scale only axis,
xmode=log,
xmin=0.2,
xmax=25,
xtick={1,10},
xticklabels={\empty},
xminorticks=true,
xmajorgrids,
xminorgrids,
ymin=0,
ymax=4,
ylabel={ (e) $c2$},
ymajorgrids,
name=plot5,
at=(plot4.below south west),
anchor=above north west,
legend style={draw=black,fill=white,legend cell align=left}
]
\addplot [color=black,solid,line width=1.5pt,mark=o,mark options={solid}]
  table[row sep=crcr]{0.2	1.22002826779223\\
0.3	0.754170670553974\\
0.4	0.715971419962266\\
0.6	0.701624575685298\\
0.8	0.696474222388648\\
1	0.696864917907875\\
1.2	0.696315989454753\\
1.6	0.696877510084155\\
2	0.696257161712063\\
2.5	0.696478496220135\\
3.2	0.696378101023417\\
4	0.695881545572635\\
5	0.696258205838101\\
7	0.696929447291716\\
10	0.695899178281394\\
15	0.695933012703295\\
20	0.696812735773523\\
25	0.696047037349452\\
};
\addlegendentry{ode-based-solver};

\addplot [color=red,solid,line width=1.5pt,mark=o,mark options={solid}]
  table[row sep=crcr]{0.2	3.67074932116648\\
0.3	1.9153197321639\\
0.4	1.20269808239569\\
0.6	0.642792404026901\\
0.8	0.457918703098305\\
1	0.335142117236033\\
1.2	0.250103280216432\\
1.6	0.158053585005525\\
2	0.109928153747862\\
2.5	0.0853241453454087\\
3.2	0.0596595404577622\\
4	0.0397921721071264\\
5	0.0268868050348108\\
7	0.0088213559578612\\
10	0.00234088960896833\\
15	1.27840790884438e-15\\
20	1.27840790884438e-15\\
25	1.27840790884438e-15\\
};
\addlegendentry{qpOASES};

\end{axis}

\begin{axis}[%
width=\fwidth,
height=\fheight,
scale only axis,
xmode=log,
xmin=0.2,
xmax=25,
xminorticks=true,
xlabel={Normalized perfomance $\bar{P}$},
xmajorgrids,
xminorgrids,
ymode=log,
ymin=1161102080.81392,
ymax=25322140497.9672,
ytick={     1000000000,     10000000000,    100000000000,   1000000000000,  10000000000000, 100000000000000},
yminorticks=true,
ylabel={ (f) $\mathcal{J}^{BF}$},
ymajorgrids,
yminorgrids,
at=(plot5.below south west),
anchor=above north west
]
\addplot [color=black,solid,line width=1.5pt,mark=o,mark options={solid},forget plot]
  table[row sep=crcr]{0.2	7685821602.72723\\
0.3	3662822281.11918\\
0.4	2685033319.69806\\
0.6	2374851365.35982\\
0.8	2353795465.89545\\
1	2352387127.80596\\
1.2	2345625390.57119\\
1.6	2347497305.92959\\
2	2344460104.9624\\
2.5	2347820882.97723\\
3.2	2348222230.4113\\
4	2342479961.28814\\
5	2344215033.33947\\
7	2349349135.78811\\
10	2342333985.18682\\
15	2342528364.60143\\
20	2348286599.07652\\
25	2343219347.81618\\
};
\addplot [color=red,solid,line width=1.5pt,mark=o,mark options={solid},forget plot]
  table[row sep=crcr]{0.2	12661070248.9836\\
0.3	7610857464.3958\\
0.4	6121046286.17612\\
0.6	3874103703.24008\\
0.8	3024768223.45011\\
1	2726724945.90996\\
1.2	2545464073.8169\\
1.6	2430057066.55885\\
2	2349279117.2791\\
2.5	2369942887.50546\\
3.2	2340495891.18508\\
4	2334331051.63474\\
5	2340183366.64594\\
7	2334426258.76709\\
10	2322204161.62784\\
15	2322513153.41887\\
20	2326341125.84746\\
25	2323012559.01895\\
};
\end{axis}
\end{tikzpicture}%
\caption{Performance indicators of the two solvers comparison vs the normalized computation power. The case $\bar P=1$ corresponds to the PLC we dispose of and which is presented in section \ref{PLCDES}.}
\label{solv_cmp}
\end{figure}
Then, constraints respect is presented in two different manners:
\begin{equation}\label{cst_resp_bo_1}
c_1 = \max_{k\in \{1,\dots,N_{sim}\}}\max_{j\in \{1,\dots,n_c\}}\max\{\Gamma_jp_k-\gamma_j,0\}\\
\end{equation}
being the maximum predicted constraints violation during the simulation while
\begin{equation}\label{cst_resp_bo_2}
c_2 = \dfrac{1}{N_{sim}} \sum _{k=1}^{N_{sim}} \max_{j\in \{1,\dots,n_c\}}max\{\Gamma_jp_k-\gamma_j,0\}
\end{equation}
being the average predicted constraints violation during the simulation.\ \\ \ \\ 
Finally, a closed-loop cost has been calculated according to: 
\begin{equation} \label{cst_sum_bf_1}
\bar{\mathcal{J}}^{BF} = \sum _{k=0}^{N_{sim}} \bar{J}^{inst}_k
\end{equation}
\ \\
The quantities (\ref{cost_sum_bo_dev}), (\ref{cost_sum_bo_cst}), (\ref{cost_sum_bo_dev})+(\ref{cost_sum_bo_cst}), (\ref{cst_resp_bo_1}), (\ref{cst_resp_bo_2}) and (\ref{cst_sum_bf_1}) are shown in Fig. \ref{solv_cmp} against  normalized computational performance $\bar{P}$. It can be noticed that the suboptimal ODE-based solver is behaving better than qpOASES in the case of low performance computation devices, while the qpOASES solver becomes clearly better beyond some hardware performance indicator.\\
The trajectories of the two closed-loop results are shown in Fig. \ref{mult_solv}, comparing the two solvers for the nominal PLC performance $P_0$ against the result obtained with the qpOASES solver with limited number of iterations and with the $10$ maximum number of iterations. It comes clearly that the use of the less efficient (per iteration) solver with $20$ iterations outperform the use of $10$ iterations of the qpOASES solver. Moreover, the use of the ODE-based solver enables the nominal qpOASES (without limitation) performance to be recovered.

\setlength\fheight{1.75cm}

\subsection{Control updating period monitoring}
In the section, attention is focused on the ODE-based solver. First of all, simulations will be done for updating period from one to five (i.e. a number of iterations from 4 to 20), and it will be shown that quadratic performances vary and there is an optimum to be found. Then the algorithm described in \cite{Alamir_ECC2013} will be implemented to show its efficiency on the cryogenic plant. \\

\begin{figure*}
\begin{center}
\input{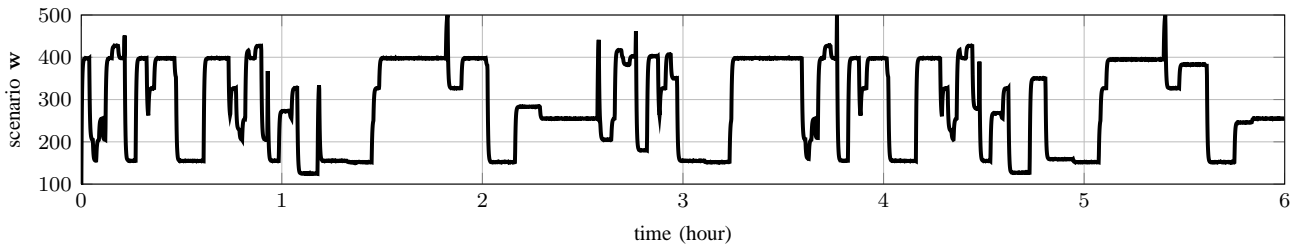}
\caption{Six hours heat loads scenario}
\label{scenarios}
\end{center}
\end{figure*}

A six hour heat loads scenario presented by Fig. \ref{scenarios} will be divided in six one hour parts, to be simulated. Cost \eqref{cost_sum_bo_dev} + \eqref{cost_sum_bo_cst} defined in the previous section will be plotted against the chosen updating period. The result is presented by Fig. \ref{tu_opt}. It can be noted that the optimum updating period is different reading the scenario. It illustrates the fact that the updating period should be monitored to enhance performance. 

The Fig. \ref{tu_opt} also plots the obtained performance by monitoring the updating period using the algorithm \cite{Alamir_ECC2013}. It can be seen that it could lead to enhanced performances.

\setlength\fheight{1.75cm}
\setlength\fwidth{3.25cm}
\begin{figure}
\begin{center}
% This file was created by matlab2tikz v0.4.7 running on MATLAB 7.14.
% Copyright (c) 2008--2014, Nico SchlÃ¶mer <nico.schloemer@gmail.com>
% All rights reserved.
% Minimal pgfplots version: 1.3
% 
% The latest updates can be retrieved from
%   http://www.mathworks.com/matlabcentral/fileexchange/22022-matlab2tikz
% where you can also make suggestions and rate matlab2tikz.
% 
\begin{tikzpicture}

\begin{axis}[%
width=\fwidth,
height=\fheight,
scale only axis,
xmin=1,
xmax=5,
xtick={\empty},
xmajorgrids,
ymin=80.3748514374999,
ymax=100,
ylabel={(c)},
ymajorgrids,
name=plot3
]
\addplot [color=black,solid,line width=1.5pt,mark=o,mark options={solid},forget plot]
  table[row sep=crcr]{1	100\\
1.5	86.4552516555444\\
2	82.3748514374999\\
2.5	85.8389125535697\\
3	88.6430398438477\\
3.5	88.1808580406351\\
4	87.9125952182724\\
4.5	91.2954546782965\\
5	98.4453813124302\\
};
\addplot [color=black,dash pattern=on 1pt off 3pt on 3pt off 3pt,line width=1.5pt,forget plot]
  table[row sep=crcr]{1	83.9061238714048\\
1.5	83.9061238714048\\
2	83.9061238714048\\
2.5	83.9061238714048\\
3	83.9061238714048\\
3.5	83.9061238714048\\
4	83.9061238714048\\
4.5	83.9061238714048\\
5	83.9061238714048\\
};
\end{axis}

\begin{axis}[%
width=\fwidth,
height=\fheight,
scale only axis,
xmin=1,
xmax=5,
xtick={\empty},
xmajorgrids,
ymin=69.747195722877,
ymax=100,
ylabel={(a)},
ymajorgrids,
name=plot1,
at=(plot3.above north west),
anchor=below south west
]
\addplot [color=black,solid,line width=1.5pt,mark=o,mark options={solid},forget plot]
  table[row sep=crcr]{1	100\\
1.5	71.747195722877\\
2	89.4537019785867\\
2.5	76.851756586193\\
3	76.2646522769686\\
3.5	72.1078205058136\\
4	86.5807298869112\\
4.5	81.0437425083514\\
5	78.597346623857\\
};
\addplot [color=black,dash pattern=on 1pt off 3pt on 3pt off 3pt,line width=1.5pt,forget plot]
  table[row sep=crcr]{1	78.6042839418278\\
1.5	78.6042839418278\\
2	78.6042839418278\\
2.5	78.6042839418278\\
3	78.6042839418278\\
3.5	78.6042839418278\\
4	78.6042839418278\\
4.5	78.6042839418278\\
5	78.6042839418278\\
};
\end{axis}

\begin{axis}[%
width=\fwidth,
height=\fheight,
scale only axis,
xmin=1,
xmax=5,
xtick={\empty},
xmajorgrids,
ymin=91.7980995685311,
ymax=100,
ylabel={(b)},
ymajorgrids,
name=plot2,
at=(plot1.right of south east),
anchor=left of south west
]
\addplot [color=black,solid,line width=1.5pt,mark=o,mark options={solid},forget plot]
  table[row sep=crcr]{1	99.7138490791599\\
1.5	98.6108236499638\\
2	97.9003084513155\\
2.5	97.0175248166198\\
3	97.2476902730761\\
3.5	98.403037235942\\
4	98.4615431180713\\
4.5	98.6019971109881\\
5	100\\
};
\addplot [color=black,dash pattern=on 1pt off 3pt on 3pt off 3pt,line width=1.5pt,forget plot]
  table[row sep=crcr]{1	93.7980995685311\\
1.5	93.7980995685311\\
2	93.7980995685311\\
2.5	93.7980995685311\\
3	93.7980995685311\\
3.5	93.7980995685311\\
4	93.7980995685311\\
4.5	93.7980995685311\\
5	93.7980995685311\\
};
\end{axis}

\begin{axis}[%
width=\fwidth,
height=\fheight,
scale only axis,
xmin=1,
xmax=5,
xtick={\empty},
xmajorgrids,
ymin=73.4992362751914,
ymax=100,
ylabel={(d)},
ymajorgrids,
name=plot4,
at=(plot2.below south west),
anchor=above north west
]
\addplot [color=black,solid,line width=1.5pt,mark=o,mark options={solid},forget plot]
  table[row sep=crcr]{1	100\\
1.5	83.0831082944143\\
2	75.8759008704281\\
2.5	77.1270037008701\\
3	75.4992362751914\\
3.5	87.0712982817508\\
4	84.2415849560746\\
4.5	87.0462031655742\\
5	96.2604574753104\\
};
\addplot [color=black,dash pattern=on 1pt off 3pt on 3pt off 3pt,line width=1.5pt,forget plot]
  table[row sep=crcr]{1	83.0263571589508\\
1.5	83.0263571589508\\
2	83.0263571589508\\
2.5	83.0263571589508\\
3	83.0263571589508\\
3.5	83.0263571589508\\
4	83.0263571589508\\
4.5	83.0263571589508\\
5	83.0263571589508\\
};
\end{axis}

\begin{axis}[%
width=\fwidth,
height=\fheight,
scale only axis,
xmin=1,
xmax=5,
xtick={1, 2, 3, 4, 5},
xlabel={updating perdiod $\tau_u$},
xmajorgrids,
ymin=81.538929386638,
ymax=100,
ylabel={(f)},
ymajorgrids,
name=plot6,
at=(plot4.below south west),
anchor=above north west
]
\addplot [color=black,solid,line width=1.5pt,mark=o,mark options={solid},forget plot]
  table[row sep=crcr]{1	100\\
1.5	86.6993964588653\\
2	83.538929386638\\
2.5	87.208934138821\\
3	87.0137679717288\\
3.5	87.8218847293525\\
4	87.4880247677863\\
4.5	89.4820802450014\\
5	90.9702665650221\\
};
\addplot [color=black,dash pattern=on 1pt off 3pt on 3pt off 3pt,line width=1.5pt,forget plot]
  table[row sep=crcr]{1	85.5196473593095\\
1.5	85.5196473593095\\
2	85.5196473593095\\
2.5	85.5196473593095\\
3	85.5196473593095\\
3.5	85.5196473593095\\
4	85.5196473593095\\
4.5	85.5196473593095\\
5	85.5196473593095\\
};
\end{axis}

\begin{axis}[%
width=\fwidth,
height=\fheight,
scale only axis,
xmin=1,
xmax=5,
xtick={1, 2, 3, 4, 5},
xlabel={updating perdiod $\tau_u$},
xmajorgrids,
ymin=63.8051923136333,
ymax=100,
ylabel={(e)},
ymajorgrids,
at=(plot6.left of south west),
anchor=right of south east
]
\addplot [color=black,solid,line width=1.5pt,mark=o,mark options={solid},forget plot]
  table[row sep=crcr]{1	100\\
1.5	65.8051923136333\\
2	69.1313129937248\\
2.5	66.7567077559635\\
3	68.4027112853062\\
3.5	69.25844944325\\
4	70.8258513589577\\
4.5	72.9728572303037\\
5	73.6733010090968\\
};
\addplot [color=black,dash pattern=on 1pt off 3pt on 3pt off 3pt,line width=1.5pt,forget plot]
  table[row sep=crcr]{1	69.1864673069146\\
1.5	69.1864673069146\\
2	69.1864673069146\\
2.5	69.1864673069146\\
3	69.1864673069146\\
3.5	69.1864673069146\\
4	69.1864673069146\\
4.5	69.1864673069146\\
5	69.1864673069146\\
};
\end{axis}
\end{tikzpicture}%
\caption{Normalized cost \eqref{cost_sum_bo_dev}+\eqref{cost_sum_bo_cst} against updating period (consequently the number of iterations) for six different scenarios named (a) to (f). Solid lines represent the cost while dotted lines depict the obtained costs with the algorithm described in \cite{Alamir_ECC2013} for $\delta =2$ }
\label{tu_opt}
\end{center}
\end{figure}
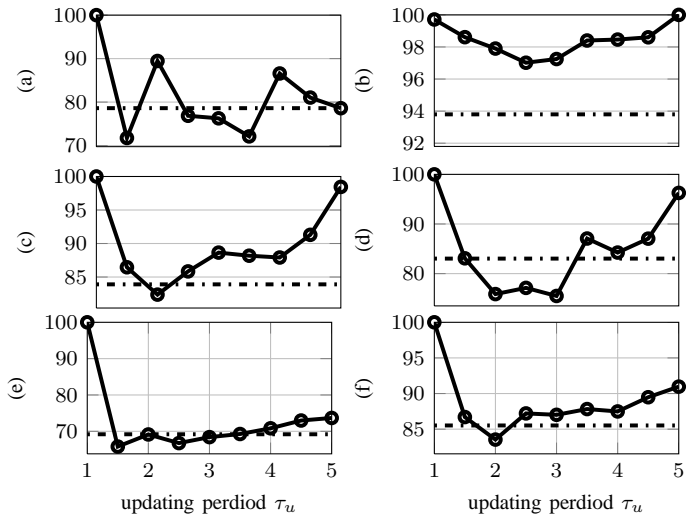

\setlength\fwidth{16cm}
\setlength\fheight{2.25cm}

\vskip 0.5cm 
\section{Experimental results} \label{secexp} 

The control scheme derived in section \ref{Theproblem} and the solver depicted in section \ref{odebasedsolver} has been implemented in the Schneider PLC described in section \ref{PLCDES}, in structured language. The objective of the section is triple. First, we want to show that the problem we derive in section \ref{Theproblem} is relevant regarding the control of a cryo-refrigerator submitted to transient heat loads.  Then, we want to emphasize that the algorithm described in \ref{odebasedsolver} is PLC compliant, event with polyhedral constraints. Finally, we will see that monitoring the updating scheme is very useful in this particular cases. 

\subsection{Control result with real time PLC implementation}
\begin{figure}
\begin{center}
\input{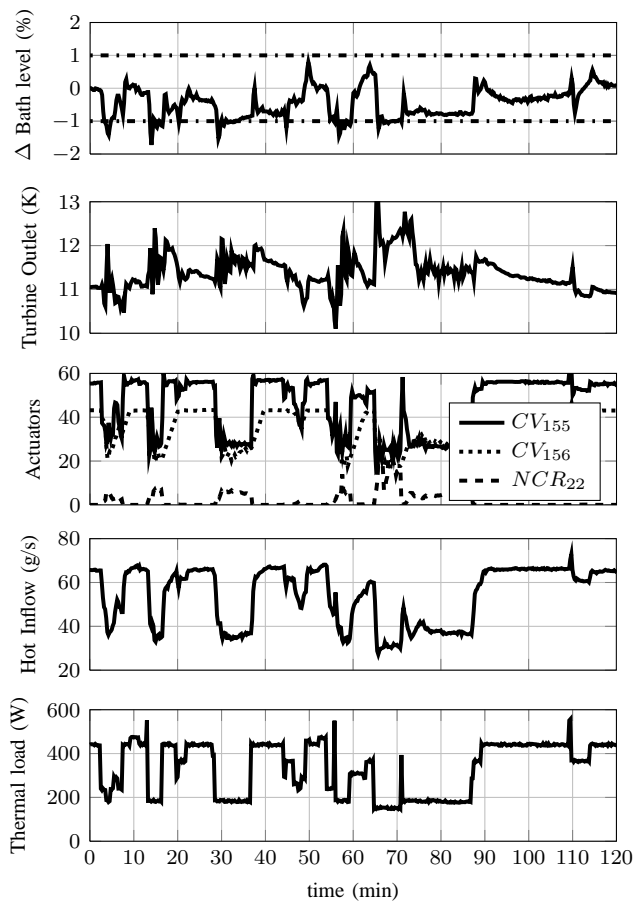}
\caption{Two hours heat load scenario. This Figure shows that the problem derived in section \ref{Theproblem} is relevant to control the plant. The $\Delta$ level represent the helium level $L1$ variation in the tank, Turbine stand for the output turbine temperature $T_5$. The Inflow depict the high pressure flow $M_{12}$ coming in the cold-box}\label{scenario_exp}
\end{center} 
\end{figure}

The plant has been submitted to a two hours scenario (first two hours of Fig. \ref{scenarios}), starting from the equilibrium. The observed time per iteration is never longer than $500ms$ as expected and the problem preparation time do not exceed $500ms$ also. It allows the optimisation algorithm to iterate $4\tau_u -1$ time.  For the first test, we chose to use a $\tau_u = 5s$ updating period. Fig. \ref{scenario_exp} shows that the control scheme is able to stabilize the plant and make the constraints to be respected, even if the pant is submitted to transient variable loads.

\setlength\fwidth{7cm}
\setlength\fheight{1.75cm}

\subsection{Some leads on the updating scheme efficiency}

\begin{figure}
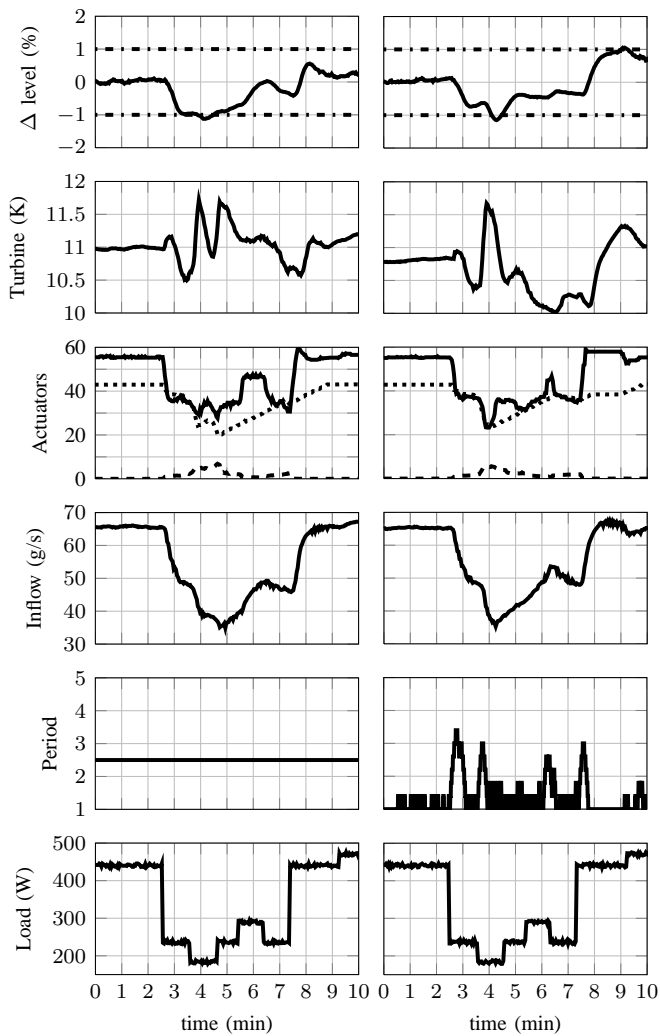

\begin{center}
\input{res_exp_tu.tikz}
 \hspace{-2cm} \hfill
\pgfplotsset{ yticklabel style = {color = red, opacity = 0.0}}
\input{res_exp_tu_fix.tikz}
\caption{Result with both constant ($2.5s$) and real-time updated updating period. It an be notice that in the case of a heat load disturbance, the updating period is increasing (in order to make the number of iteration to also increase), since the hot-starting solution is far from the actual solution. Period represent the updating period $\tau_u$. For actuators legend, please refer to Fig. \ref{scenario_exp}. }
\label{diff_upt}
\end{center}
\end{figure}

The algorithm to update the updating period as been implemented on the PLC to show its efficiency. Unfortunately, the cost is not monitored but it is still possible to show result in the time domain.  Fig \ref{diff_upt} shows the difference between a constant updating period and a variable one. One can see that in the case of a serious change on the thermal load, the updating period is increasing to iterate more, while the algorithm is imposing a short updating period as soon as the problem is not changing much from an updating instant to another.

\setlength\fwidth{3.5cm}
\setlength\fheight{1.75cm}

\vskip 0.5cm 
\section{Conclusion} \label{secconc} 
In this paper, an efficient way to control a cryogenic plant submitted to variable heat loads using an industrial PLC with reduced computing capabilities is proposed. It has been shown that in this application, an ODE-based solver gives robust sub optimal solution even in the case where the hot-start is far from the optimal solution (in the case of an unpredicted disturbance for instance). The control scheme and the solver has been both validated experimentally.
 
Moreover an algorithm that automatically monitors the control updating period has been implemented and experimentally successfully tested.
 
Future investigation will aim at developing an MPC control scheme for a refrigerator submitted to multiple different thermal loads, including at $1.8K$  (super-fluid helium). Also, cryogenic systems could be very large (several buildings): the control scheme will be distributed in order to ensure a progressive integration to industrial systems.

\vskip 0.5cm 
\section*{Acknowledgment}

Authors would like to thank every co-worker from the SBT for their kind help to improve models and control
strategy and for their time to correct and discuss this paper. Authors give special thanks to Michel Bon-Mardion, Lionel
Monteiro, François Millet, Christine Hoa, Bernard Rousset and Jean-Marc Poncet from SBT for their explanation about
the process and their participation on experimental campaigns.

\ifCLASSOPTIONcaptionsoff
  \newpage
\fi

\bibliographystyle{plain}
\bibliography{bibBonnecryo}

\begin{thebibliography}{10}

\bibitem{Alamir2006}
M.~Alamir.
\newblock {\em Stabilization of nonlinear systems using receding-horizon
  control schemes: A parameterized approach for fast systems}.
\newblock Springer-Verlag, 2006.

\bibitem{Alamir_pavia2008}
M.~Alamir.
\newblock A framework for monitoring control updating period in real-time
  {NMPC} schemes.
\newblock In Lalo Magni, DavideMartino Raimondo, and Frank Allgöwer, editors,
  {\em Nonlinear Model Predictive Control}, volume 384 of {\em Lecture Notes in
  Control and Information Sciences}, pages 433--445. Springer Berlin
  Heidelberg, 2009.

\bibitem{Alamir_ECC2013}
M.~Alamir.
\newblock Monitoring control updating period in fast gradient based {NMPC}.
\newblock In {\em Control Conference (ECC), 2013 European}, pages 3621--3626,
  July 2013.

\bibitem{Alamir_ECC2014}
M.~Alamir.
\newblock Fast {NMPC}, a reality-steered paradigm: Key properties of fast
  {NMPC} algorithms.
\newblock In {\em Control Conference (ECC), 2014 European}, June 2014.

\bibitem{trbdf2}
R.~E. Bank, William M.~Coughran Jr., Wolfgang Fichtner, Eric Grosse, Donald~J.
  Rose, and R.~Kent Smith.
\newblock Transient simulation of silicon devices and circuits.
\newblock {\em IEEE Trans. on CAD of Integrated Circuits and Systems}, pages
  436--451, 1985.

\bibitem{Bemporad2012}
A.~Bomporad and P.~Patrinos.
\newblock Simple and certifiable quadratic programming algorithms for embedded
  linear model predictive control.
\newblock In {\em Proceeding of the IFAC Nonlinear Predictive Control
  Conference}, Noordwijkerhout, NL, 2012.

\bibitem{bonne_cec_2}
F.~Bonne, M.~Alamir, and P.~Bonnay.
\newblock Physical control oriented model of large scale refrigerators to
  synthesize advanced control schemes. design, validation, and first control
  results.
\newblock In {\em Proceedings of the Cryogenic Engineering Conference}, 2013.

\bibitem{bonne_cec_1}
F.~Bonne, M.~Alamir, P.~Bonnay, and B.~Bradu.
\newblock Model based multivariable controller for large scale compression
  stations. design and experimental validation on the lhc 18kw
  cryorefrigerator.
\newblock In {\em Proceedings of the Cryogenic Engineering Conference}, 2013.

\bibitem{bonne:hal-00922066}
Fran{\c c}ois Bonne, Mazen Alamir, and Patrick Bonnay.
\newblock {Nonlinear observers of the thermal loads applied to the helium bath
  of a cryogenic Joule-Thompson Cycle}.
\newblock {\em Journal of Process Control}, 24(3):73--80, January 2014.

\bibitem{clavel_these}
F.~Clavel.
\newblock {\em Mod\'{e}lisation et contr\^ole d'un r\'{e}frig\'{e}rateur
  cryog\'{e}nique. Application \`a la station 800W \`a 4.5K du CEA Grenoble}.
\newblock PhD thesis, EEATS, 2011.

\bibitem{clavel}
F.~Clavel, M.~Alamir, P.~Bonnay, A.~Barraud, G.~Bornard, and C.~Deschildre.
\newblock Multivariable control architecture for a cryogenic test facility
  under high pulsed loads: Model derivation, control design and experimental
  validation.
\newblock {\em Journal of Process Control}, 21(7):1030--1039, 2011.

\bibitem{diehl2005real}
M.~Diehl, H.~G. Bock, and J.~P. Schl{\"o}der.
\newblock A real-time iteration scheme for nonlinear optimization in optimal
  feedback control.
\newblock {\em SIAM Journal on control and optimization}, 43(5):1714--1736,
  2005.

\bibitem{Diehl2005IEE}
M.~Diehl, R.~Findeisen, F.~Allgower, H.~G. Bock, and J.~P. Schloder.
\newblock Nominal stability of real-time iteration scheme for nonlinear model
  predictive control.
\newblock {\em IEE Proc.-Control Theory Appl.}, 152(3):296--308, 2005.

\bibitem{Ferreau2008}
H.J. Ferreau, H.G. Bock, and M.~Diehl.
\newblock An online active set strategy to overcome the limitations of explicit
  mpc.
\newblock {\em International Journal of Robust and Nonlinear Control},
  18(8):816--830, 2008.

\bibitem{Ferreau2014}
H.J. Ferreau, C.~Kirches, A.~Potschka, H.G. Bock, and M.~Diehl.
\newblock {qpOASES}: A parametric active-set algorithm for quadratic
  programming.
\newblock {\em Mathematical Programming Computation}, 2014.
\newblock (in print).

\bibitem{Jones2012}
C.~N. Jones, A.~Domahidi, M.~Morari, S.~Richter, F.~Ullmann, and M.~Zeilinger.
\newblock Fast predictive control: Real-time computation and certification.
\newblock In {\em Proceeding of the IFAC Nonlinear Predictive Control
  Conference}, Noordwijkerhout, NL, 2012.

\bibitem{Mayne2000}
D.~Q. Mayne, J.~B. Rawlings, C.~V. Rao, and P.~O.~M. Scokaert.
\newblock Constrained model predictive control: Stability and optimality.
\newblock {\em Automatica}, 36:789--814, 2000.

\bibitem{Nesterov1983}
Y.~Nesterov.
\newblock A method of solving a convex programming problem with convergence
  rate o (1/k2).
\newblock {\em Soviet Mathematics Doklady}, 27(2):372--376, 1983.

\bibitem{doc_plc}
Schneider Electric, avalaible at http://www.schneider-electric.com/products/.
\newblock {\em Schneider PLC TSXP574634M Product data sheet}.

\bibitem{Zavala2008}
V.~M. Zavala, C.~D. Laird, and L.~T. Biegler.
\newblock Fast implementation and rigorous models: can both be accommodated in
  {NMPC}?
\newblock {\em International Journal of Robust and Nonlinear Control},
  18(8):800--815, 2008.

\end{thebibliography}

\end{document}